\documentclass[aps,pra,twocolumn,superscriptaddress]{revtex4}
\usepackage{graphicx}
\usepackage{times}
\usepackage{bm}
\usepackage{amsmath,amssymb}
\usepackage{feynmf}
\usepackage{multirow}
\usepackage[colorlinks=true, allcolors=blue]{hyperref}
\usepackage{xcolor}
\usepackage{array}
\usepackage{siunitx}
\usepackage{tabularx}
\usepackage{mathtools}
\usepackage{esint}

\begin{document}


\title{Multipole decomposition of the thermal one-loop self-energy correction for a bound atomic electron
}

\author{J. J. Lopez-Rodriguez}
\affiliation{ 
Department of Physics, St. Petersburg State University, Petrodvorets, Oulianovskaya 1, 198504, St. Petersburg, Russia
}
\author{A. Bobylev}
\affiliation{ 
Department of Physics, St. Petersburg State University, Petrodvorets, Oulianovskaya 1, 198504, St. Petersburg, Russia
}
\affiliation{ 
Petersburg Nuclear Physics Institute named by B.P. Konstantinov of National Research Centre 'Kurchatov Institut', St. Petersburg, Gatchina 188300, Russia
}
\author{P. Kvasov}
\affiliation{ 
Department of Physics, St. Petersburg State University, Petrodvorets, Oulianovskaya 1, 198504, St. Petersburg, Russia
}

\author{T. Zalialiutdinov}
\affiliation{ 
Department of Physics, St. Petersburg State University, Petrodvorets, Oulianovskaya 1, 198504, St. Petersburg, Russia
}
\affiliation{ 
Petersburg Nuclear Physics Institute named by B.P. Konstantinov of National Research Centre 'Kurchatov Institut', St. Petersburg, Gatchina 188300, Russia
}
\author{D. Solovyev}
\email[E-mail:]{d.solovyev@spbu.ru}
\affiliation{ 
Department of Physics, St. Petersburg State University, Petrodvorets, Oulianovskaya 1, 198504, St. Petersburg, Russia
}
\affiliation{ 
Petersburg Nuclear Physics Institute named by B.P. Konstantinov of National Research Centre 'Kurchatov Institut', St. Petersburg, Gatchina 188300, Russia 
}

\begin{abstract}
In this paper, we present a comprehensive analysis of the one-loop self-energy correction at finite temperature for the bound electron. In this approach, we study the influence of thermal radiation on atomic systems. Along the way, we found well-known effects, including thermal Stark and Zeeman shifts, as well as thermal quadrupole interactions and relativistic corrections to the multipole expansion of photon field operators. We show that the corresponding contributions arise from the decomposition of the fully relativistic expression in terms of the $\alpha Z$ parameter. The presented analysis unambiguously determines the consistency of the quantum electrodynamics theory at finite temperature (TQED) with the perturbation theory of quantum mechanics (QM). Although our analysis mainly focuses on the hydrogen atom model, their potential implications for precision spectroscopic experiments are discussed.

\end{abstract}

\maketitle


\section{Introduction}
\label{intro}

It is well-known that blackbody radiation (BBR) leads to energy shifts and line broadening in atomic systems \cite{PhysRevLett.42.835,PhysRevA.23.2397}. With the increasing experimental precision in recent decades, BBR-induced energy shifts have become particularly important for modern atomic spectroscopy. Specifically, experiments measuring the hyperfine splitting (HFS) of the ground state for cesium, rubidium, and strontium atoms, used as atomic clocks, have been conducted with uncertainties ranging from $2.3 \times 10^{-16}$ to $2. 1 \times 10^{-18}$ \cite{Nicholson_2015, Pizzocaro_2017}, while for the hydrogen atom the HFS interval of the ground state has been measured with a relative experimental uncertainty of about $1.7 \times 10^{-12}$ \cite{Hellwig}. There are many other examples of precision measurements of transition frequencies in various atomic systems, see, for example, \cite{van_Rooij_2011, Matveev}. 
At the accuracy level mentioned above, temperature-dependent energy shifts (induced by external thermal radiation) act as one of the main sources of error in determining frequencies, and therefore require careful control of the thermal environment in such experiments. 
Although this leads to a significant complication of the experimental setup, in this case the measurement result can be consolidated with high accuracy theoretical calculations, thus providing a way to further increase the accuracy \cite{Itano,Stark-measurement,Beloy}. The achieved accuracy and its corresponding increase should facilitate the verification of increasingly subtle effects, the precise determination of various physical quantities, the resolution of existing discrepancies between theory and experimental results, etc. 

As the best-known phenomenon caused by blackbody radiation, studies of the ac Stark shift (and the dc Stark shift as its most significant contribution) are widely reported in the literature for various atomic and molecular systems. Since the thermal Stark effect is defined by the dipole polarizability of the atom, at leading order such a shift is independent of the hyperfine quantum numbers and thus does not contribute to HFS measurements \cite{Safronova2010}. As a consequence, the search for the influence of the thermal environment on hyperfine transitions is reduced to next order effects and, in particular, to the thermal magnetic shift (thermal Zeeman shift) \cite{Itano,Han}. 

Within the framework of quantum mechanics, the black-body Stark shift (BBRS) and the Zeeman shift (BBRZ) can be described by second-order perturbation theory. In this context, the quadratic perturbation interaction reduces to Planck's radiation law, which can be related to electric and magnetic fields by the following expressions \cite{PhysRevA.23.2397,Itano} (in relativistic units):
\begin{eqnarray}
\label{1}
   \mathcal{E}^2(\omega) = \mathcal{B}^2(\omega) =  \frac{8}{\pi}\frac{\omega^3}{e^{\beta \omega}-1},
\end{eqnarray}
together with $\beta=1/k_{\rm B}T$ ($k_{\rm B}$ is the Boltzmann constant and $T$ is the radiation temperature in Kelvin). Performing integration over frequency in Eq.~(\ref{1}) and going to SI units, the mean square fields can be found \cite{riehle2006frequency}:
\begin{eqnarray}
\langle \mathcal{E}^2(t)\rangle = \frac{1}{2}
\int\limits_0^\infty \mathcal{E}^2(\omega)d\omega = (831.9 \text{ V/m})^2\left[\frac{\text{T}(\text{K})}{300}\right]^4,
\label{2}
\\
\langle \mathcal{B}^2(t)\rangle = 
(2.775 \times 10^{-6} \text{ Tesla})^2\left[\frac{\text{T}(\text{K})}{300}\right]^4.
\label{3}
\end{eqnarray}

In most cases the thermally induced Zeeman shift is much smaller than the Stark shift. For the transition frequencies between low-lying atomic states at room temperature, the Stark shift is typically of the order of a few hertz, see \cite{PhysRevA.23.2397}. In contrast, for transitions between hyperfine sublevels of the same level in atoms and ions with nonzero nuclear spin, the leading order Stark shift is absent, while the Zeeman shift becomes dominant (the same holds for the fine splitting). In particular, for microwave atomic clocks, the fractional BBRZ shift to the working transition is typically of the order of $10^{-17}$ \cite{Han}. This fact imposes severe limits on the stability of the clocks \cite{riehle2006frequency}.


Within QED theory, thermal energy shifts and BBR-induced level widths for the bound electron can be obtained by modifying the photon propagator \cite{PhysRevA.92.022508}. A derivation of the finite-temperature contribution to the particle propagation functions can be found, e.g., in \cite{Dolan:1973qd,PhysRevD.28.340,DONOGHUE1985233}. In the case of atomic systems and laboratory conditions, the finite-temperature modification of the fermion propagator is irrelevant.  
As in the case of zero temperature \cite{LabKlim}, within the framework of finite-temperature QED theory (TQED), the corresponding Stark shift and induced level width arise from the real and imaginary parts of the one-loop self-energy correction in the leading order, respectively.

As will be shown in this paper, in order to obtain the atomic level energy shift due to thermal magnetic interaction in the TQED approach, it is necessary to evaluate the next order in parameter $\alpha Z$ relativistic corrections. Taking into account laboratory conditions, we focus mainly on the energy shifts and BBR-induced widths at room temperature $T = 300$ K, implying its application to spectroscopic experiments.

For the sake of consistency, the paper is organised as follows. First, the non-relativistic derivation of the thermal Stark shift within ordinary QM second-order perturbation theory is briefly discussed in section~\ref{qm}. Then the corresponding fully relativistic approach within the TQED formalism is presented in section~\ref{se}. The various forms of multipolar decomposition are also discussed in section~\ref{se}. The analytical derivation of the blackbody radiation induced BBRZ shift is given in section~\ref{Zs}. The section~\ref{theory} continues with the derivation of relativistic corrections, as well as the diamagnetic contribution, to thermally induced shifts and the incorporation of finite-lifetime atomic level effects. Numerical results for the real and imaginary parts of the thermal one-loop self-energy correction are discussed in section~\ref{numbers}. Concluding remarks are given in the last section, and additional calculation details are given in the appendices. Throughout this paper, unless otherwise stated, relativistic units (r.u.) are used, where $\hbar = c = m_{e}=1$ ($c$ is the speed of light, $m_{e}$ is the mass of the electron, and $\hbar$ is the reduced Planck constant). For clarity, the electron charge $e$ and in some places the mass of an electron $m_{e}$ is written explicitly. We denote nonrelativistic (Schr\"odinger) energies by $E$ and relativistic (Dirac) energies by $\varepsilon $. The corresponding functions are represented using circular and triangular bra and ket vectors, respectively.

\section{Theory}
\label{theory}
\subsection{The thermal Stark shift: nonrelativstic QM approach}
\label{qm}

To derive the Stark shift stimulated by blackbody radiation within the framework of the QM approach, one should consider the second order perturbation theory \cite{PhysRevA.23.2397}. Noting that the magnetic components of the BBR are much smaller than the electric ones, we can neglect any effects that may be induced by the magnetic field. In this section we restrict our estimation of the atomic level energy shifts to the electric field induced by the BBR, see Eq.~(\ref{2}). The lower order Zeeman thermal shift can be obtained in a similar way by replacing the electric dipole interaction by the corresponding magnetic operator. 

The Hamiltonian of the atomic system exposed to the external homogeneous electric field can be written as
\begin{eqnarray}\label{Hamiltonian}
    \hat{H}= \hat{H}_0 + \hat{V},
\end{eqnarray}
where $\hat{H}_0$ denotes the unperturbed nonrelativistic Hamiltonian of the atom, and $\hat{V}$ represents the external perturbation, which in the dipole approximation is equal to $e\bm{r} \bm{\mathcal{E}}$. Here $e$ is the electron charge, $\bm{r}$ is the radius vector of the electron and $\bm{\mathcal{E}}$ denotes the static external electric field.

According to the selection rules \cite{Sob}, the diagonal matrix element of the dipole moment is zero: $( nlm_l|\hat{V}|nlm_l ) = 0$. Here $n$ is the principal quantum number, $l$ is the orbital momentum and $m_l$ is its projection. Since we are dealing with the same orbital momentum $l$, and the dipole interaction "mixes" states of opposite parity, this circumstance is valid for all non-degenerate energy states in the first-order correction (the linear Stark effect is absent). 

In second-order perturbation theory, the general form of the quadratic Stark effect is given by
\begin{eqnarray}
\label{quadraticstark}
    \Delta E_{a}  = \sum \limits_{\substack{n \ne a}} \frac{|( a|\hat{V}| n )|^2}{E_a-E_n}     
= e^2 \sum_{\substack{i,\,n \ne a}} \frac{\mathcal{E}^2_{i} |( a|r_{i}| n  )  |^2}{E_a-E_n},
\end{eqnarray}
where the sum over $n$ implies the summation over the entire spectrum of Schr\"odinger equation, including continuum, and $\mathcal{E}_{i}$ ($i=1,\,2,\,3$) denotes the component of the vector $\bm{\mathcal{E}}$. Note that the state $n=a$ does not occur according to the second-order perturbation theory.

The dynamic (ac-) Stark shift caused by interaction with an alternating electric field $\bm{\mathcal{E}}(t) = \bm{\mathcal{E}}_0 \cos(\omega t)$ can be calculated from the static response of the atom to the root mean square (RMS) electric field strength, see \cite{riehle2006frequency}, as follows 
\begin{eqnarray}\label{nonresonantshift}
    \Delta E_{a}  = \sum_{\substack{i,\,n \ne a}}\sum\limits_{\pm} \frac{e^2 \mathcal{E}_{0i}^2}{4} \frac{|( a|r_{i}| n )|^2}{E_a-E_n \pm \omega},
\end{eqnarray}
where the summation $\pm$ means the sum of two contributions with $+\omega$ and $-\omega$ in denominators. 

If the radiation source has a certain spectrum, i.e. $\mathcal{E}_{0i}=\mathcal{E}_{0i}(\omega)$ in Eq.~(\ref{nonresonantshift}), the total dynamic Stark shift is given by the integration of Eq.~(\ref{nonresonantshift}) over frequency \cite{PhysRevA.23.2397}. 
Then, taking into account that the blackbody radiation is isotropic, i.e. $\mathcal{E}_{0i}^2(\omega) = \frac{1}{3}\mathcal{E}^2(\omega)$, the expression for the dynamic Stark shift reduces to
\begin{eqnarray}\label{farley'sformula}
    \Delta E_{a}^{\mathrm{BBRS}}= \frac{2e^2}{3\pi} \text{P.V.}\int\limits_{0}^{\infty} \frac{d\omega\,\omega ^3}{ e^{\beta \omega}-1 } 
   \sum\limits_{n \neq a}   \sum\limits_{\pm}
    \frac{\left |  ( a|  \bm{r} |n )  \right|^2}{E_a-E_n \pm \omega}.
\end{eqnarray}
Here, any singularities in the integrals, generated by resonant frequencies, are taken in terms of the principal value, referred to as $\text{P.V.}$

\subsection{One-loop self-energy correction for the bound electron: Stark shift and line broadening}
\label{se}

\begin{figure}[ht] 
\centering
\includegraphics[width=0.7\columnwidth]{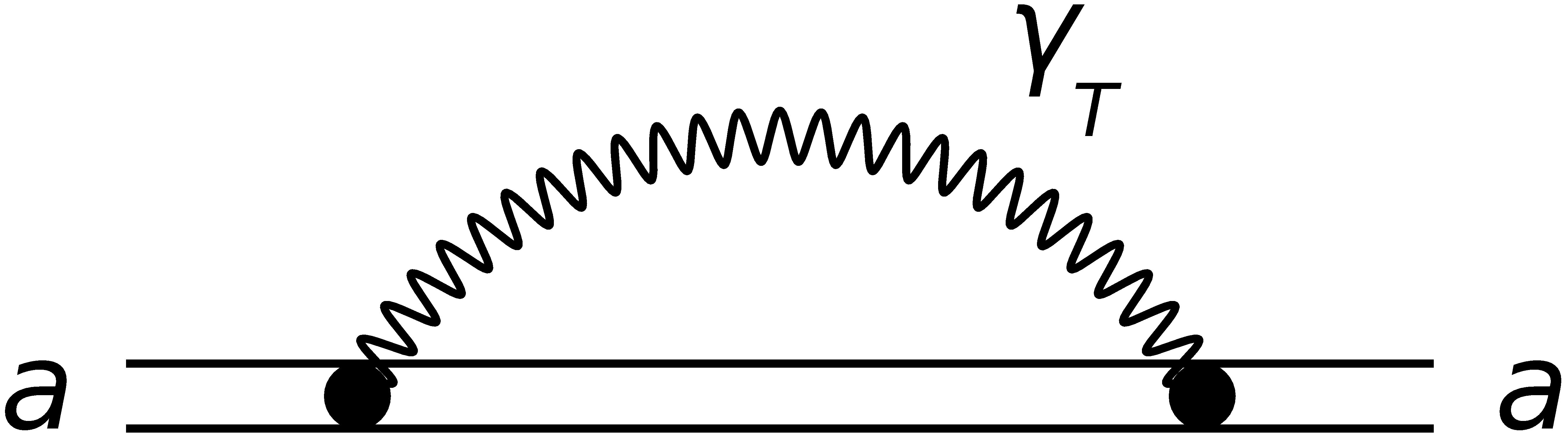} 
\caption{Thermal one-loop self-energy correction for the energy of an atomic electron at level $a$. The double solid line denotes a bound electron in the Furry picture, the part between two vertices (indicated by filled circles) sets the electron propagator. The inner wavy line denotes the virtual photon described either by the ordinary photon propagator or by the thermal one, the latter is emphasized by the index $\gamma_T$. }
\label{fig:1}
\end{figure}

In the previous section, the thermally induced Stark shift was derived in the framework of QM theory. To obtain a similar energy shift using the QED approach, a thermal one-loop self-energy correction for the bound electron was analyzed in \cite{PhysRevA.92.022508}. The Feynman diagram of the one-loop self-energy correction for the bound electron is shown in Fig.~\ref{fig:1}.

The radiative correction given by the Feynman graph in Fig.~\ref{fig:1} can be evaluated within the $S$-matrix formalism. The corresponding matrix element is
\begin{eqnarray}
\label{S-matrix}
   \langle f |\hat{S}^{(2)}| i \rangle= (-ie)^2 \int d^4 x_1 d^4 x_2 \overline{\psi}_f(x_1) \gamma^{\mu} S(x_1,x_2)\times
\\
\nonumber   
    D_{\mu \nu}(x_1,x_2)\gamma^{\nu} \psi_i(x_2),
\end{eqnarray}
where $\gamma^{\mu}$ are the Dirac matrices, $\mu=0,\,1,\,2,\,3$. The space-time vector $x=(t,\bm{r})$, where $t$ is the time variable and $\bm{r}$ is the spacial radius-vector as before. The Dirac electron wave function, $\psi_{a}(x)$, for an arbitrary state $a$ can be written as $\psi_{a}(x)= \psi_{a} (\bm{r}) e^{-i \varepsilon_{a} t}$, and $\overline{\psi}$ means a Dirac conjugated wave function. 

The terms $S(x_1,x_2)$ and $D(x_1,x_2)$ are the electron and photon propagators, respectively. The electron propagator can be taken in the standard form \cite{Akhiezer}:
\begin{eqnarray}
    \label{el_prop}
    S(x_1,x_2)=\frac{i}{2\pi}\int\limits_{-\infty}^{+\infty}d\Omega\, e^{-i\Omega(t_1-t_2)}
    \sum\limits_{n}\frac{\psi_{n}(\bm{r}_1)\overline{\psi}_{n}(\bm{r}_2)}{\Omega - \varepsilon_{n} (1-i0)}\,,
\end{eqnarray}
whereas the photon function $ D_{\mu_1 \mu_2}(x_1,x_2)$ consists of two contributions \cite{PhysRevA.92.022508}:
\begin{eqnarray}
\label{phpr}
D_{\mu_1 \mu_2}(x_1,x_2) = \frac{g_{\mu_1 \mu_2}}{2\pi \mathrm{i} r_{12}}\int\limits_{-\infty}^{+\infty}d\omega e^{\mathrm{i}|\omega|r_{12} - \mathrm{i}\omega(t_1-t_2)} 
\\
\nonumber
- \frac{g_{\mu_1 \mu_2}}{\pi r_{12}}\int\limits_{-\infty}^{+\infty}d\omega\, n_{\beta}(|\omega|) \sin{|\omega|r_{12}} \, e^{-\mathrm{i}\omega(t_1-t_2)}.
\end{eqnarray}
Here the function $n_{\beta}(\omega)$ is defined by the Planck distribution, $n_{\beta}(\omega)= \left(\mathrm{exp}\left (\omega/k_{\rm B} T\right)-1\right)^{-1}$, $g_{\mu\nu}$ is the metric tensor, and $r_{12}\equiv|\bm{r}_1-\bm{r}_2|$. The existence of the sum of two contributions in Eq.~(\ref{phpr}) allows one to search for thermal corrections arising from the Planck distribution independently of the first (ordinary) one. Further calculations of Eq.~(\ref{S-matrix}) in the zero temperature case can be found in \cite{LabKlim}, where it is shown that the real part represents the contribution to the Lamb shift and the imaginary part gives the width of the excited state.

Focusing on the finite temperature case and omitting intermediate calculations for brevity (see \cite{PhysRevA.92.022508,SOLOVYEV2020168128} for details), the expression for the energy shift can be obtained as
\begin{eqnarray}\label{Energy shift}
 \Delta E_a = \frac{e^2}{\pi} \sum_n \left[  \frac{1-\bm{\alpha}_1\bm{\alpha}_2}{r_{12}} I^{\beta}_{na}(r_{12})\right]_{anna}.
\end{eqnarray}
Here the matrix element $[\dots]_{anna}$ should be understood as $[\hat{A}(12)]_{abcd}\equiv \langle a(1) b(2)|\hat{A} | c(1) d(2)\rangle$, $\bm{\alpha}$ - Dirac matrix, and
\begin{eqnarray}
\label{Ir}
I^{\beta}_{na}(r_{12}) = \int\limits^{\infty}_0 d\omega\, n_{\beta}(\omega) \sum\limits_{\pm}\frac{\sin\omega r_{12}}{\varepsilon_a-\varepsilon_n \pm \omega + \mathrm{i}\,0}.
\end{eqnarray}

Using known partial decomposition
\begin{eqnarray}
    \frac{\sin \omega r_{12}}{\omega r_{12}} = \sum\limits_{L=0}^{\infty}(2L+1)j_{L}(\omega r_{1})j_{L}(\omega r_{2})
    \\\nonumber\times
    \bm{C}_{L}(\bm{n}_1)\bm{C}_{L}(\bm{n}_2),
\end{eqnarray}
where $\bm{n}=\bm{r}/|\bm{r}|$,  $j_{L}(\omega r)$ is the spherical Bessel function of the first kind \cite{GradRyzh} and $\bm{C}_{L}$ are vector spherical harmonics \cite{varsh}, the thermal shift takes the form:
\begin{eqnarray}
    \label{multipole}
\Delta E_a = \frac{e^2}{\pi} \sum\limits_{L} (2L + 1)\int\limits_{0}^{\infty}d\omega\,\omega\,n_{\beta}(\omega)
\\\nonumber
\times\sum\limits_{n}
\sum\limits_{\pm}\frac{\langle a |\alpha_{\mu}j_{L}(\omega r)\bm{C}_{L}|n\rangle \langle n| \alpha^{\mu}j_{L}(\omega r)\bm{C}_{L}|a \rangle}{\varepsilon_a-\varepsilon_n \pm \omega + \mathrm{i}\,0}
\\\nonumber
=
\frac{e^2}{\pi} \sum\limits_{L} (2L + 1)\int\limits_{0}^{\infty}d\omega\,\omega\,n_{\beta}(\omega)
\alpha_{L}(\omega).
\end{eqnarray}

In the last line of Eq.~(\ref{multipole}) $\alpha_{L}(\omega)$ represents the relativistic generalization of the $2^{L}$-pole dynamic polarizability \cite{ach2010}. It should be noted that, unlike the similar partial-wave decomposition of the ordinary bound-state self-energy (see Eq.~(12.87) in the textbook \cite{Lindgren2011}), the expression~(\ref{multipole}) has no ultraviolet divergences due to the presence of the cut-off factor $n_{\beta}(\omega)$ in the integrand. In addition to the expression (\ref{multipole}), the original Eq.~(\ref{Energy shift}) can also be expanded in terms of multipole operators of the photon field and thus express the thermal shift in terms of the relativistic tensor of linear scattering \cite{PhysRevA.74.020502}.  

With the use of Sokhotski–Plemelj theorem, $I^{\beta}_{na}(r_{12})$ can be written as
\begin{eqnarray}\label{Ir12}
 I^{\beta}_{na}(r_{12}) = \mathrm{P.V.}\int\limits^{\infty}_0 d\omega\, n_{\beta}(\omega) \sum\limits_{\pm}\frac{\sin\omega r_{12}}{\varepsilon_a-\varepsilon_n \pm \omega} + 
 \\
 \nonumber
  + \mathrm{i} \pi \int\limits^{\infty}_0 d\omega\, n_{\beta}(\omega) \sin(\omega r_{12}) \sum\limits_{\pm} \delta(\varepsilon_n -\varepsilon_a \pm \omega).
\end{eqnarray}
Below we omit the notation P.V., assuming that all integrations over frequency in the real part are taken as the principal value.

Substituting Eq.~(\ref{Ir12}) into Eq.~(\ref{Energy shift}) we get an expression consisting of real and imaginary parts. Afterwards in the nonrelativistic limit the dipole approximation at $\omega r \sim (\alpha Z) {\rm\, r.u.} \ll 1$ can be utilized. Then $\sin\omega r_{12}$ can be decomposed into a Taylor series. Leaving aside the first three terms in Eq.~(\ref{Energy shift}) and considering that $r^2_{12}=r^2_1+r^2_2-2(\bm{r}_1\bm{ r}_2)$, the expression for the one-electron atom, see \cite{PhysRevA.92.022508} for details, is
\begin{eqnarray}\label{dE}
\Delta E_a = \frac{4e^2}{3 \pi} \sum_{n} | ( a |\bm{r} | n )|^2 \int\limits^{\infty}_0 d\omega\, n_{\beta}(\omega) 
\frac{\omega_{na}\omega^3}{\omega^2_{na} - \omega^2}  
\\
\nonumber
- \frac{2  e^2}{3} \mathrm{i} \sum_{n } n_{\beta}(|\omega_{an}|)\,|\omega_{an}|^3\, | ( a |\bm{r} | n )|^2,
\end{eqnarray}
in conjunction with the notation $\omega_{na}=E_n -E_a$. It is noteworthy that, in the transition to the nonrelativistic limit in the velocity form, the contribution of the negative continuum in the summation over the entire Dirac spectrum in Eq.~(\ref{Energy shift}) in the leading order collapses into a constant, which subsequently vanishes upon conversion to the length form (see Appendix~\ref{dia} for some details).

The real part of Eq.~(\ref{dE}) is exactly the same as the nonrelativistic expression~(\ref{farley'sformula}) for the dynamical Stark shift. The parametric estimate of the $\Delta E_a^{\mathrm{BBRS}}$ shift in hydrogen-like ions with nuclear charge $Z$ can be given as follows. Taking into account that in relativistic units $\omega\sim m_{e}(\alpha Z)^2$, $r\sim (m_{e}\alpha Z)^{-1}$ and $\int_{0}^{\infty}\omega^3n_{\beta}(\omega) \sim (k_{B}T)^4$ we find that for the ground state $\Delta E_{a}^{\mathrm{BBRS}} \sim \frac{(k_{B}T)^4}{\alpha^3 m_{e}^3 Z^4}$ r.u., which is in agreement with the estimation given in \cite{PhysRevA.78.042504}. 

The imaginary part of Eq.~(\ref{dE}) gives rise to the thermally induced level width (depopulation rate) according to the relation \cite{LabKlim}:
\begin{eqnarray}\label{level width}
\Gamma_a =  -2{\rm Im} \Delta E_a
\end{eqnarray}
In other words, the expression for the self-energy of the bound electron obtained in the TQED theory exactly reproduces the quantum mechanical result in the leading order. Note that the summation in the imaginary part includes all states, including those above a given arbitrary atomic level $a$ \cite{PhysRevA.23.2397}. This difference from the natural width (where only the lower states are summed) is due to the possibility of excitation in an external field. In particular, the summation over upper states leads to a non-zero width of the ground state of the atom \cite{PhysRevA.78.042504}.

\subsection{One-loop self-energy correction for the bound electron: Zeeman shift and line broadening}
\label{Zs}

The expression (\ref{dE}) was derived in the nonrelativistic limit, where the dipole matrix elements are calculated using Schr\"odinger wave functions. A phenomenological approach can be employed to account for the magnetic interaction by
replacing the electric dipole operator in Eq.~(\ref{dE}) by the magnetic one. This is the standard way to account
for BBRZ shift in hyperfine transitions \cite{Itano,Han}. However, the primary objective of the present study is to derive the thermal Zeeman shift within the framework of QED theory at finite temperature.

As mentioned in section~\ref{intro}, the magnetic components of the BBR cause a much smaller effect than the electric components (according to Maxwell's equations, the magnetic field strength is suppressed by the factor $1/c$). At the same time, as shown above, the ac-Stark shift and the level width result from the one-loop self-energy correction for the bound electron in the leading order. It is therefore worth looking at the relativistic corrections discarded in Eq.~(\ref{dE}).

Returning to the Taylor series expansion of $\sin$ in Eq.~(\ref{Ir12}), the leading-order relativistic correction is defined by
\begin{eqnarray}\label{relc}
\Delta E_a^{\rm rel}= \frac{e^2 }{6 \pi} \sum\limits_n \int\limits^{\infty}_0 d\omega\, \omega^3\,n_{\beta}(\omega)  \langle  a n| (\bm{\alpha}_1 \bm{\alpha}_2) r^2_{12} 
  | n a \rangle \,\,\,
  \\
  \nonumber
\times\left( \frac{1}{\varepsilon_a-\varepsilon_n + \omega + \mathrm{i}\, 0} +  \frac{1}{\varepsilon_a-\varepsilon_n - \omega + \mathrm{i}\,0}   \right).
\end{eqnarray}
The summands in the expansion of $\sin\omega r_{12}  $ containing the fifth power of the frequency and higher are discarded, as they contribute additional smallness due to the extra temperature factors $\beta^{-2}$.

The expression for $\Delta E_a^{\rm rel}$ can be simplified using the relation $r^2_{12}=r^2_1+r^2_2-2\bm{r}_1\bm{r}_2$, which allows us to separate the magnetic polarizability, given that $(\bm{\alpha}_1 \bm{\alpha}_2) (\bm{r}_1\bm{r}_2) = ([\bm{r}_1\times\bm{\alpha}_1]\cdot[\bm{r}_2\times\bm{\alpha}_2]) + (\bm{r}_1\bm{\alpha}_2)(\bm{r}_2\bm{\alpha}_1)$. For details on the derivation of the thermal Zeeman shift, as well as the corresponding derivation for the thermal quadrupole interaction, see Appendix~\ref{ApB}. 

Then, the expression~(\ref{relc}) can be rewritten as follows
\begin{eqnarray}
\label{rel}
\Delta E_a^{\rm rel} = \frac{e^2}{6\pi}\sum\limits_n \int\limits^{\infty}_0 d\omega\, \omega^3\,n_{\beta}(\omega) 
\langle  a n| 
(\bm{\alpha}_1 \bm{\alpha}_2)(r_1^2+r_2^2) 
\nonumber
\\
- 2([\bm{r}_1\times\bm{\alpha}_1]\cdot[\bm{r}_2\times\bm{\alpha}_2]) - 2(\bm{r}_1\bm{\alpha}_2)(\bm{r}_2\bm{\alpha}_1)| na \rangle \qquad
\\
\nonumber
\times\left( \frac{1}{\varepsilon_a - \varepsilon_n + \omega + \mathrm{i}\,0} +  \frac{1}{\varepsilon_a - \varepsilon_n - \omega + \mathrm{i}\, 0}   \right).\qquad
\end{eqnarray}
Introducing the operator of magnetic moment $\bm{\mu} = e[\bm{r}\times\bm{\alpha}]/2$ \cite{PhysRevA.56.R2499}, we find
\begin{eqnarray}
\label{relm}
\Delta E_a^{\rm rel} = -\frac{2}{3\pi}\sum\limits_{n} \int\limits^{\infty}_0 d\omega\, \omega^3\,n_{\beta}(\omega) 
\Big[ \langle  a n| (\bm{\mu}_1 \bm{\mu}_2)| na \rangle 
\nonumber
\\
-\frac{e^2}{4}\langle  a n| (\bm{\alpha}_1 \bm{\alpha}_2)(r_1^2+r_2^2) - (\bm{\alpha}_1\bm{\alpha}_2)(\bm{r}_1\bm{r}_1) | na \rangle + \qquad
\\
\nonumber
 \frac{e^2}{4}\langle  a n| (\bm{r}_1\bm{\alpha}_2)(\bm{r}_2\bm{\alpha}_1)| na \rangle  \Big]
\sum\limits_{\pm}
\frac{1}{\varepsilon_a-\varepsilon_n \pm \omega + \mathrm{i}\,0},\qquad
\end{eqnarray}
where the magnetic interaction term is emphasized separately. 

In the nonrelativistic limit (see Appendix~\ref{ApBadd}), the electron magnetic moment, $\bm{\mu}$, is given by
\begin{eqnarray}
\label{mu}
   \bm{\mu}= - \mu_{B} (\bm{l}+2\bm{s}),
\end{eqnarray}
where $\bm{s}=\bm{\sigma}/2$ is the electron spin momentum, $\bm{\sigma}$ is the Pauli matrix, $\bm{l}$ is the orbital momentum ($\bm{j}=\bm{l}+\bm{s}$ is the total angular momentum operator) and $\mu_{B} = |e|/2m_e$ is the Bohr magneton. Passing to the nonrelativistic limit in the matrix elements for the first term in Eq.~(\ref{relm}), one can obtain
\begin{eqnarray}
\label{Zshift}
  \Delta E_a^{\mathrm{}} = -\frac{2}{3\pi}
\sum\limits_n ( a n|   (\bm{\mu}_1\bm{\mu}_2) |n a ) \int\limits^{\infty}_0 d\omega\, n_{\beta}(\omega)\,\omega^3 
  \\
  \nonumber
  \times
  \sum\limits_{\pm} \frac{1}{E_a-E_n \pm \omega + i 0} .
\end{eqnarray}
The calculation of the matrix element $\langle a| (\bm{l} + 2 \bm{s}) |n \rangle$ can be performed in a fully analytical way using the Wigner-Eckart theorem \cite{varsh}.

Similarly as in the previous subsection~\ref{se}, the real and imaginary parts in Eq.~(\ref{Zshift}) can also be distinguished: \( \Delta E_a = \Delta E^{\rm BBRZ}_a - \mathrm{i} \Gamma_a^\beta(\mathrm{M}1)/2 \). The real contribution is the expected thermal Zeeman shift (BBRZ). The imaginary contribution corresponds to the induced width of the atomic level formed by the sum of magnetic dipole transitions, denoted as M1. Since the magnetic interaction of the electron with the external field is expressed by the spin and orbital moments Eq.~(\ref{mu}), only states of the same parity as level $a$ survive in the sum over intermediate states, resulting in a non-zero effect for transitions between fine or hyperfine sublevels of the atom. 

Considering transitions corresponding to a fine or hyperfine atomic structure, due to the smallness of the splitting interval, one can expect a "strengthening"\, of the effect. Furthermore, it becomes increasingly relevant for BBRZ that the energy denominator can approach zero in the case of resonance, $\omega \sim \omega_{an}$. The theory for regularizing the resonance contribution is well established. Within the framework of QED theory, this procedure was introduced in \cite{Low}  and discussed in more detail in \cite{Andr}. Regularization in the context of the Stark thermal shift was explored in \cite{PhysRevA.92.022508}, while its application to black-body friction enhancement was presented in \cite{PhysRevLett.108.043005}. For the sake of self-consistency, the corresponding analyses are detailed in the next section. Here we restrict ourselves to the result in Eq.~(\ref{Zshift}) to align it with the well-known quantum mechanics expression, as discussed in \cite{Itano,PhysRevA.110.043108}.

According to \cite{Itano}, the leading contribution to the Zeeman thermal shift comes from the nearest state $a'$ of the same parity as $a$ in the spectral sum in Eq.~(\ref{Zshift}). Hence, 
\begin{eqnarray}
\label{BBRZ}
  \Delta E^{\rm BBRZ}_a = -\frac{4}{3\pi}  |(  a| \bm{\mu} |a' ) |^2
\int\limits^{\infty}_0 d\omega  \frac{n_{\beta}(\omega)\omega^3 \omega_{aa'}}{\omega_{aa'}^2-\omega^2}.
\end{eqnarray}
The expression (\ref{BBRZ}), given in a nonrelativistic form and admitting an obviously relativistic representation, see Eq.~(\ref{relm}), completely coincides with the result known from the quantum mechanics approach. 

In turn, the imaginary part of $\Delta E_{a}$ in Eq.~(\ref{Zshift}) with the use of Eq.~(\ref{level width}) reads as follows
\begin{eqnarray}
\label{M1rt}
W_{aa'}^{\mathrm{ind}}(\mathrm{M1}) = \frac{4}{3}n_\beta(|\omega_{aa'}|)\omega_{aa'}^3 |(  a| \bm{\mu} |a' ) |^2.
\end{eqnarray}
Here we have replaced the level width by the partial transition probability with the emission of a magnetic dipole photon, since only the state $a'$ with the same parity as $a$ survives in the leading order. The expression (\ref{M1rt}) represents the well-known QM result for the induced transition probability. The factor $n_\beta(|\omega_{aa'}|)$ can be regarded as an enhancement factor (in the case $>1$) or as leading to a thermal correction (in the case $<1$) of the spontaneous transition rate. The transition probabilities caused by blackbody radiation dominate over the spontaneous ones at frequencies $\omega_{aa'}\leq k_{\rm B} T$ and vice versa for $k_{\rm B} T\ll \omega_{aa'}$.

It can be shown that the remaining summands in Eq.~(\ref{relm}) give rise to the thermal quadrupole interaction as well as relativistic corrections to the electric dipole interaction. The corresponding derivations are given in Appendix~\ref{ApB}. The final expression for the quadrupole thermal shift in the nonrelativistic limit can be written as follows
\begin{eqnarray}
\label{BBRQ}
\Delta E_a^{\mathcal{Q}} = - \frac{e^2}{90\pi}\sum\limits_n\int\limits_0^\infty d\omega\, 
\frac{\omega_{an} \omega^5 n_\beta}{\omega_{an}^2-\omega^2}
(an| \mathcal{Q}^{(1)}_{ij} \mathcal{Q}^{(2)}_{ij} |na)\,\,\,
\end{eqnarray}
with $\mathcal{Q}^{(1)}_{ij} = 3r_{1i} r_{1j}-r^2_1\delta_{ij}$. In turn, the remaining relativistic correction simplifies 
\begin{eqnarray}
\label{BBRr}
\Delta E_a^{\rm rem} =   
\frac{e^2}{3\pi}\frac{\pi^4(k_{\rm B} T)^4}{15}
( a |  r^2(\bm{r}\cdot \bm{\nabla})| a),
\end{eqnarray}
where integration over frequency was performed, resulting in a $T^4$ dependence.

\subsection{One-loop self-energy correction for the bound electron: impact of finite life-times of atomic levels}
\label{reg}

\begin{figure}[ht] 
\centering
\includegraphics[width=0.8\columnwidth]{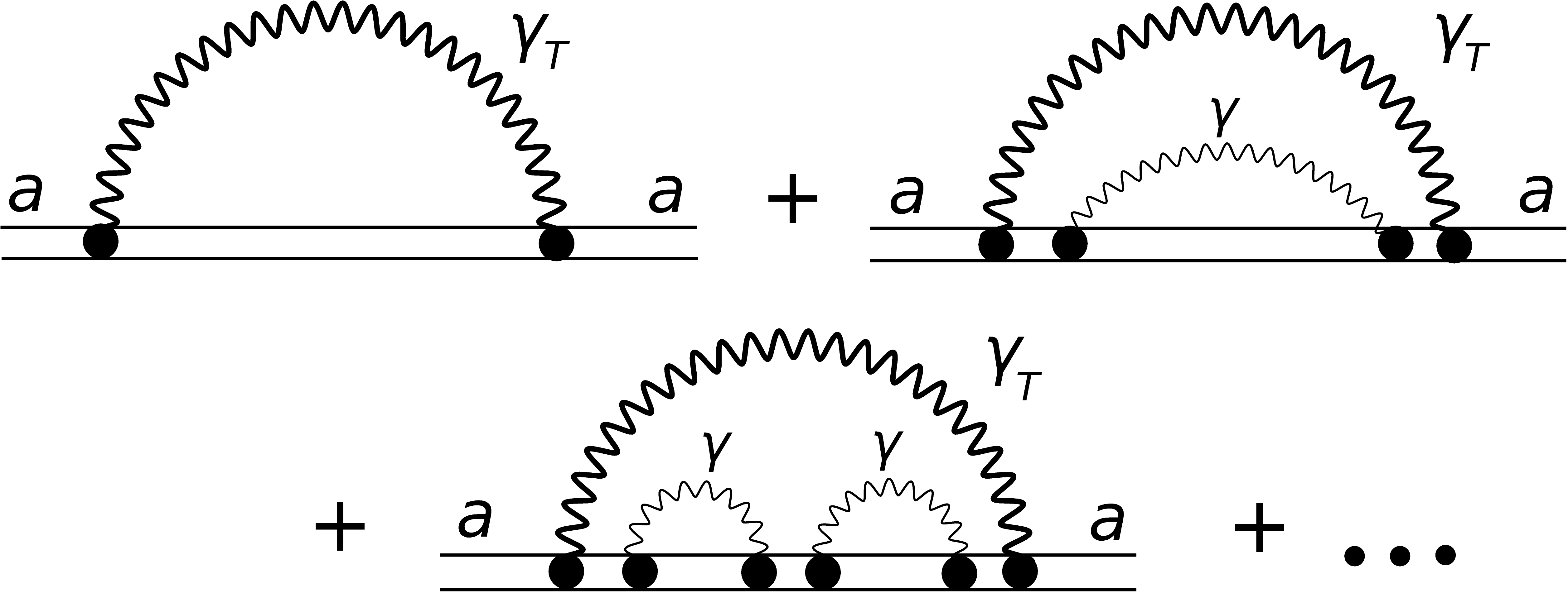} 
\caption{Insertions of ordinary (zero-temperature) self-energy loops into the thermal self-energy correction, see Fig.~\ref{fig:1}, for an arbitrary atomic state $a$. The zero temperature loop is labelled $\gamma$, while the finite temperature loop is labelled $\gamma_T$.}
\label{fig:2}
\end{figure}

The expression~(\ref{BBRZ}) was obtained by expanding $\sin$ contained in the thermal self-energy QED operator using the Sokhotski-Plemelj formula. The latter can be considered as a quantum mechanics approach.
In this section, we consider a special case where the energy denominator in Eq.~(\ref{Zshift}) becomes singular. The QED regularization procedure consists of including the usual zero-temperature one-loop self-energy corrections in the corresponding $S$-matrix element Eq.~(\ref{S-matrix}), see \cite{Low,Andr,ZSL_PhysRep2018}. In this case, one should evaluate an infinite set of Feynman diagrams as shown in Fig.~\ref{fig:2}: inside the thermal loop, the usual zero-temperature self-energy corrections are progressively inserted - in the order of loop after loop \cite{PhysRevA.92.022508}.

An infinite series of successive insertions of zero-temperature self-energy loops leads to a geometric progression, the final result of which is the appearance of the self-energy correction of the electron in the diverging denominator. The real part of this correction represents the energy shift (dominant part of the Lamb shift) that can effectively be included in the energy $\varepsilon_a$, and the imaginary part gives half of the natural level width \cite{LabKlim}. The presence of an imaginary additive in the divergent energy denominator leads to a regular expression (Lorentz contour in the case of photon scattering processes \cite{ZSL_PhysRep2018}). The details of such a derivation have recently been given in \cite{PhysRevA.92.022508}, leading to a noticeable change in the BBR-induced level widths. However, the Stark shift remains almost unchanged when the finite lifetimes of the levels are taken into account \cite{PhysRevA.78.042504,PhysRevA.92.022508}.

The enhancement effect in a blackbody environment due to the finite lifetime of the atomic levels has been known since \cite{PhysRevLett.108.043005}. For the thermomagnetic interaction one would expect a similar behavior as for the Stark shift. However, there is a notable difference between the Stark effect and the BBRZ effect. It is related to the frequencies involved in the calculations, see the analysis in \cite{Han,PhysRevA.110.043108}. Resonant contributions to the thermal Stark shift are suppressed by the size of the Planck distribution. For atomic frequencies corresponding to fine and hyperfine transitions, $n_\beta(\omega)$ can be several orders of magnitude higher due to the overlap of the Planck distribution function and the resonance energy. Therefore, considering finite lifetimes of atomic levels can lead to a significant change in the final value of the Zeeman thermal shift. A similar effect was recently observed for the BBR frictional forces in \cite{PhysRevLett.108.043005}.

The QED regularization is much simpler for the BBRZ shift in the context of numerical calculations because there is no summation over intermediate states in the expression (\ref{BBRZ}). 
Note that in the case of two excited states the imaginary additive is represented by the sum of the widths \cite{Andr,ZSL_PhysRep2018}. Thus, for the regularized BBRZ shift one can obtain
\begin{eqnarray}
\label{BBRZ-2}
  \Delta E^{\rm BBRZr}_a = -\frac{2}{3\pi}  |( a| \bm{\mu} |a' ) |^2
\int\limits^{\infty}_0 d\omega\,\omega^3\,n_{\beta}(\omega) \times
\\
\nonumber
\left( \frac{\omega_{aa'}+\omega}{(\omega_{aa'}+\omega)^2+\frac{1}{4}\Gamma^2} + \frac{\omega_{aa'}-\omega}{(\omega_{aa'}-\omega)^2+\frac{1}{4}\Gamma^2}\right),
\end{eqnarray}
where it is assumed that $\omega_{aa'}$ can be positive and negative, and $\Gamma=\Gamma_a+\Gamma_{a'}$ in the case of two excited states. 

\subsection{Relativistic corrections to wave function}
\label{corr}

Above, the derivation of the thermal Stark and Zeeman shifts from the one-loop self-energy correction for the bound electron was presented. The expressions are given in the nonrelativistic limit with the corresponding contributions arising from the expansion of the thermal self-energy operator. However, relativistic corrections to the wave function should also be taken into account. For the thermal Stark effect, these corrections are of the same order of magnitude as the Zeeman thermal shift.

The leading correction to the large component of the Dirac wave function can be written in the form \cite{LabKlim,JentAdkins}:
\begin{eqnarray}\label{Lc1}
\varphi_{a}= \left[  1 - \frac{(\bm{\sigma} \bm{p})^2}{8 m_e^2c^2}  \right] \phi_{a},
\end{eqnarray}
where $\phi_{a}$ satisfies the nonrelativistic Schr\"odinger equation, and the second term is a correction of order $\alpha^2$ (the speed of light $c$ is written explicitly for clarity). Consequently, for $Z = 1$, this correction to the thermal Stark shift is of the same order as the Zeeman shift \cite{PhysRevA.110.043108}, i.e. $\frac{(k_{B}T)^4}{\alpha m_{e}^3Z^2}$. 

The Stark thermal shift refers to the matrix element $(\tilde{a}|\bm{r}|\tilde{n})$ (see equation~(\ref{dE})), where the tilde states denote the large component of the Dirac wave functions this time. Substitution of (\ref{Lc1}) into the dipole matrix element leads to
\begin{eqnarray}
\label{Lc2}
|(\tilde{a}|\bm{r}|\tilde{n})|^2\approx \left|(a|\bm{r}|n) - \frac{1}{8m_e^2c^2}(a|p^2\bm{r}+\bm{r}p^2|n)\right|^2,
\end{eqnarray}
where the higher order terms $\sim 1/c^4$ are omitted and the wave functions in the right-hand side of the equation are assumed to be completely nonrelativistic. The second term in Eq.~(\ref{Lc2}) can be simplified using the Schr\"odinger equation: $p^2|n) = 2m_e (E_n-V)|n)$, where $V$ is the Coulomb interaction potential for the hydrogen-like atom. The contribution $\bm{r}V$ vanishes because of the orthogonality of the radial wave functions. In this case, the product $\bm{r}V\sim \bm{n}$ is a unit vector along the radius vector, and hence the corresponding matrix element results in $n=a$. Since then $\omega_{an}=0$ in the numerator of Eq.~(\ref{dE}), this contribution is zero. Thus we arrive at
\begin{eqnarray}
\label{Lc3}
|(\tilde{a}|\bm{r}|\tilde{n})|^2\approx |(a|\bm{r}|n)|^2 - \frac{E_a+E_n}{2m_e c^2}|(a|\bm{r}|n)|^2.
\end{eqnarray}

Finally, the correction to the thermal Stark shift via the wave function modification is
\begin{eqnarray}
\label{Lc4}
\Delta E_a^{\rm wf} = -\frac{2e^2}{3 \pi} \sum_{n} | ( a |\bm{r} | n )|^2 
\\
\nonumber
\times
\int\limits^{\infty}_0 d\omega\,\omega^3 n_{\beta}(\omega) 
\frac{E_{n}^2-E_a^2}{\omega^2_{na} - \omega^2}.\qquad
\end{eqnarray}
The corresponding parametric estimate for $\omega_{na}\sim m_{e}(\alpha Z)^2$ is $\Delta E_a^{\rm wf} \sim\frac{(k_{B}T)^4}{\alpha m_{e}^2Z^2}$. The numerical values of this correction for some states of the hydrogen atom are presented in the next section of the paper.

\subsection{Diamagnetic interaction owing to negative spectrum}
\label{di}

In nonrelativistic quantum mechanics the interaction of an atomic electron with an external magnetic field is specified, along with the linear term, by the square of the vector potential \cite{Landau,JentAdkins}. This contribution, called the diamagnetic interaction, is proportional to the square of the bound electron radius vector and the square of the magnetic field strength. Then, according to formula~(\ref{1}), the diamagnetic interaction should be manifested in the thermal case as well.

To derive the thermal diamagnetic energy shift we turn to the expressions~(\ref{Energy shift}) and (\ref{Ir}). The not yet examined contribution corresponds to the negative spectrum occurring in the relativistic Dirac theory. Recall that states of the negative continuum are also included in the sum over the entire spectrum in Eq.~(\ref{Energy shift}). Thus up to now only "large"\, components of the Dirac wave functions from the positive spectrum and corrections to them have been considered. Denoting the states of the negative Dirac continuum as $n^{(-)}$, and the energies respectively $\varepsilon_{n^{(-)}}$, at lowest order the energy denominator in Eq.~(\ref{Ir}) is $\varepsilon_a-\varepsilon_{n^{(-)}}\approx 2m_e c^2$. Still, the expansion of the $\sin$ into a Taylor series in the vicinity of a small argument is valid because the frequency is given by the domain of the Planck distribution (when $\omega\rightarrow m_ec^2$, $n_\beta\rightarrow 0$ exponentially). We should regard also the completeness condition, which in the nonrelativistic limit can be applied to the states of the positive and negative spectra separately. 

Accordingly, we arrive at (see Appendix~\ref{dia} for details of the derivation)
\begin{eqnarray}
\label{di.1}
 \Delta E_a^{(-)}\approx - \frac{e^2}{3\pi}\int\limits^{\infty}_0 d\omega\, n_{\beta}(\omega)\omega^3\langle a|r^2| a\rangle
\end{eqnarray}
and, therefore,
\begin{eqnarray}
\label{di.2}
\Delta E_a^{\rm dia} = \frac{\pi^3 e^2}{45\beta^4}\langle a | r^2| a\rangle,
\end{eqnarray}
where the remaining integration over frequency $\omega$ was analytically performed.

For the hydrogen atom the matrix element is $\langle a | r^2| a\rangle = \frac{n_a^2}{2}(5n_a^2+1-3l_a(l_a+1))$, where $n_a$ is the principal number and $l_a$ is the orbital momentum for the arbitrary state $a$. To obtain the magnitude of the thermal diamagnetic shift the coefficient in Eq.~(\ref{di.2}) can be expressed in terms of room temperature as
\begin{eqnarray}
\label{di.3}
\frac{\pi^3 e^2}{45\beta^4}\rightarrow 7.6425\times 10^{-8} \left[\frac{T}{300\, \text{K}}\right]^4 \text{ Hz.}
\end{eqnarray}
Thus, for the ground state in the hydrogen atom, the contribution Eq.~(\ref{di.2}) is $2.293\times 10^{-7}$ Hz and in principle is negligibly small even with respect to the relativistic corrections to the Stark shift (see the next section). However, one should note the growth of $\Delta E_a^{\rm dia}$ with increasing principal number. A significant contribution can be expected for the Rydberg states. For instance, calculations using the equations ~(\ref{di.2}), (\ref{di.3}) for $n=100,l=0$ give an energy shift $\Delta E_{n_a=100,l_a=0}^{\rm dia} = 19.1$ Hz at room temperature.

\section{Numerical Results}
\label{numbers}
\subsection{Thermal Stark shift}

In this section we present numerical results for the thermal Stark shift at different temperatures for various states of the hydrogen atom. The values obtained take into account the Lamb shift and can be compared with the results of \cite{PhysRevA.78.042504,PhysRevA.92.022508,Thermal}, where a comparative analysis with \cite{PhysRevA.23.2397} is also presented. In particular, in \cite{PhysRevA.92.022508} it was shown that the Lamb shift contribution decreases with increasing principal quantum number of the considered state $n_a$ as well as with increasing temperature. Nevertheless, for a more rigorous evaluation in the case of highly excited states, we have used in our calculations the Lamb shift values borrowed from \cite{Czarnecki2005}. Accurate estimates of the dynamical Stark shift not only confirm the presented theory, but also reveal the role of the corresponding relativistic corrections like Eq.~(\ref{BBRr}). 

The numerical results for the Stark shift (the real part of Eq.~(\ref{dE})) are collected in Table~\ref{Table1}.
\begin{table}[ht]
\centering
\caption{Thermal ac-Stark shift for $ns$ ($n=2, 3, 4, 6$) states of the hydrogen atom for different temperatures. The second line for the temperature value $T=300$ K corresponds to the values from the work \cite{PhysRevA.23.2397}. All values are given in Hz.}
\begin{tabular}{ c  c  c  c  c }
\hline
\hline
\mbox{State} & $2s$ & $3s$ & $4s$ & $6s$ \\
\hline
$77$ K& $-4.94\times10^{-3}$&$-3.80\times10^{-2}$&$-0.19$&$2.11$\\
$290$ K& $ -0.91 $  & $ -7.67  $ & $ -43.79 $    & $ -262.68 $ \\
$300$ K& $ -1.04 $  & $ -8.79  $ & $ -50.80 $    & $ -273.48 $ \\
        & $ -1.08 $  & $ -9.10  $ & $ -51.19 $    & $ -274.70  $ \\
$310$ K& $ -1.19 $  & $ -10.05 $ & $ -58.68 $    & $ -282.11 $ \\
$320$ K& $ -1.35 $  & $ -11.43 $ & $ -67.51 $    & $ -288.38 $ \\
$330$ K& $ -1.53 $  & $ -12.95 $ & $ -77.36 $    & $ -292.09 $ \\
\hline
\end{tabular}
\label{Table1}
\end{table}
All values are in good agreement with \cite{PhysRevA.23.2397}, proving the small contribution of the Lamb shift to the thermal Stark shift. We also note that the contribution of the continuous spectrum included in our calculations remains minor at the chosen temperatures and the indicated states.

The values of the correction to the Stark shift, $\Delta E_a^{\rm rem}$, due to the decomposition by the $\alpha Z$ parameter of the one-loop self-energy operator for the bound electron, Eq.~(\ref{BBRr}), are given in Table~\ref{tab:rem}.
\begin{table}[ht]
\centering
\caption{The relativistic correction, $\Delta E_a^{\rm rem}$, for different states of the hydrogen atom at different temperatures. The first and third columns show the states for which the thermal shifts Eq.~(\ref{BBRr}) are calculated. The corresponding values are given in the following columns for the temperature $300$ Kelvin. All values are in Hz.}
\begin{tabular}{ c  c  c  c  c }
\hline
\hline
\mbox{State} & $2s$ & $3s$ & $4s$ & $6s$ \\
\hline
$77$ K & $ -5.52 \times 10^{-8}$  & $-2.75 \times 10^{-7} $ & $ -8.60 \times 10^{-7}$ & $ -4.25 \times 10^{-6}$\\
$290$ K& $-1.11 \times 10^{-5}$  & $-5.52\times 10^{-5}$ & $-1.73 \times 10^{-4}$ & $-8.56 \times 10^{-4}$\\
$300$ K& $ -1.27\times 10^{-5} $  & $ -6.33\times 10^{-5}  $ & $ -1.98\times 10^{-4} $ & $-9.80 \times 10^{-4} $\\
$310$ K& $ -1.45 \times 10^{-5}$  & $ -7.21 \times 10^{-5}$ & $-2.26 \times 10^{-4}$ & $-1.12 \times 10^{-3}$\\
$320$ K& $-1.65 \times 10^{-5}$  & $-8.19 \times 10^{-5}$ & $-2.56 \times 10^{-4}$ & $-1.27 \times 10^{-3}$\\
$330$ K& $-1.86\times 10^{-5}$  & $-9.26 \times 10^{-5}$ & $-2.90 \times 10^{-4} $ & $-1.43 \times 10^{-3}$\\
\hline
\end{tabular}
\label{tab:rem}
\end{table}
The relative contribution of the correction Eq.~(\ref{BBRr}) with respect to the dynamical Stark effect for the states listed in Table~{tab:rem} remains at $10^{-5}$ for the chosen temperature range. Typically, the thermal Stark effect is considered in the static limit, for which the decomposition of the atomic polarizability by the small frequency $\omega$ is utilized. The consequence of such a decomposition consists of dynamical corrections to the dc-Stark shift. According to the results of Table~{tab:rem}, the relativistic correction $\Delta E_a^{\rm rem}$ is comparable to the dynamical corrections and, therefore, needs to be taken into account in the appropriate precision measurements.

In addition, the correction of Eq.~(\ref{BBRr}) in its form can be compared with the diamagnetic contribution Eq.~(\ref{di.2}). The $r^2$ factor included in it shows that for the Rydberg states the $\Delta E_a^{\rm rem}$ can be significant, however, still hardly exceeds the Stark effect tending to the value of a few kHz \cite{PhysRevA.23.2397}. We restrict ourselves here to calculations for the states listed in Table~{tab:rem} in order to emphasize the importance of the relativistic corrections arising from the operator decomposition into a series over $\alpha Z$. One can expect much worse experimental accuracy of measurement of transition frequencies involving Rydberg states, moreover, for transitions between Rydberg states there should be significant reductions of the correction value.

\subsection{Relativistic thermal correction Eq.~(\ref{Lc4})}

Numerical estimates for the correction arising from the wave function to the Stark thermal shift, see Eq.~(\ref{Lc4}), are presented in Table~\ref{tab:3} for some low-lying states of the hydrogen atom. The summation over the entire spectrum of Shcr\"odinger equation for hydrogen is performed with the use of B-spline approach, see \cite{Grant2009}.

\begin{table}[h]
\centering
\caption{The relativistic correction, $\Delta E_a^{\rm wf}$, to the thermal ac-Stark effect expressed by the equation (\ref{Lc4}). All values are in Hz.}
\begin{tabular}{ c  c  c  c  c }
\hline
\hline
\mbox{State} & $2s$ & $3s$ & $4s$ & $6s$ \\
\hline
$77$ K & $ 1.99 \times 10^{-8} $  & $ 8.22\times 10^{-8}$ & $2.41 \times 10^{-7} $ & $ 1.80\times 10^{-6} $\\
$290$ K & $4.02 \times 10^{-6}$  & $ 1.69 \times 10^{-5}$ & $ 5.33 \times 10^{-5}$ &  $-1.21 \times 10^{-4} $\\
$300$ K& $ 4.60\times 10^{-6} $  & $1.94\times 10^{-5}$ & $6.09\times 10^{-5}$ &  $-1.67 \times 10^{-4} $\\
$310$ K & $ 5.25 \times 10^{-6}  $  & $ 2.22 \times 10^{-5} $ & $ 6.89 \times 10^{-5}$ &  $-2.21 \times 10^{-4} $\\
$320$ K & $5.96 \times 10^{-6}$  & $2.53 \times 10^{-5}$ & $ 7.74 \times 10^{-5}$ &  $-2.83 \times 10^{-4}$\\
$330$ K & $6.74 \times 10^{-6}$  & $2.87 \times 10^{-5}$ & $8.62 \times 10^{-5}$ &  $-3.54 \times 10^{-4}$\\
\hline
\end{tabular}
\label{tab:3}
\end{table}
It can be seen from Table~\ref{tab:3} that the magnitude of the correction Eq.~(\ref{Lc4}) increases with the growth of the principal quantum number, reaching the millihertz level already for the $6s$ state. Comparing this correction $\Delta E_a^{\rm wf}$ with $\Delta E_a^{\rm rem}$, one can also see their approximately equal order, comparable to the dynamical corrections to the thermal dc-Stark shift. However, for lower states the correction from the wave function has the opposite sign.


The cumulative contributions $\Delta E_a^{\rm wf}+\Delta E_a^{\rm rem}$, as well as their relative magnitude to the ac-Stark shift $(\Delta E_a^{\rm wf}+\Delta E_a^{\rm rem})/\Delta E_a^{\rm BBRS}$ at the temperature of $300$ K are presented in Table~\ref{tab:3a}.
\begin{table}[h]
\centering
\caption{The total contributions $\Delta E_a^\Sigma=\Delta E_a^{\rm wf}+\Delta E_a^{\rm rem}$ and ac-Stark shift $\Delta E_a^{\rm BBRS}$ in Hz for the different states in the hydrogen atom, as well as their relative magnitude to the ac-Stark shift $\delta_a^\Sigma=(\Delta E_a^{\rm wf}+\Delta E_a^{\rm rem})/\Delta E_a^{\rm BBRS}$ at temperature $300$ K.}
\begin{tabular}{ c  c  c  c}
\hline
\hline
\mbox{State} & $\Delta E_a^\Sigma$, Hz & $\Delta E_a^{\rm BBRS}$, Hz & $\delta_a^\Sigma$ \\
\hline
$2s$ & $-8.1 \times 10^{-6}$ & $-1.04$ & $7.79\times 10^{-6}$ \\
$3s$ & $-4.4 \times 10^{-5}$ & $-8.79$ & $4.99\times 10^{-6}$ \\
$4s$ & $-1.4 \times 10^{-4}$ & $-50.8$ & $2.70\times 10^{-6}$ \\
$6s$ & $-1.1 \times 10^{-3}$ & $-273.48$ & $4.19\times 10^{-6}$ \\
\hline
\end{tabular}
\label{tab:3a}
\end{table}

Thus, the relative contribution of relativistic corrections to the dynamic Stark effect remains more or less at the level of $10^{-6}$. The change in its magnitude refers to the sign variation of $\Delta E_a^{\rm wf}$. It can also be argued that due to the change in sign of the main contribution $\Delta E_a^{\rm BBRS}$ depending on the atomic state, the total contribution of $\Delta E_a^\Sigma$ can also be of opposite sign. Therefore, for each atomic state under consideration, the total contribution $\Delta E_a^\Sigma$ should be calculated separately. However, this thermal effect arising from the relativistic correction to the wave function and the interaction operator is always at the level of dynamic corrections to the static Stark effect.

\subsection{The finite lifetimes effect on the Stark energy shift}
\label{lifetimes}
In view of taking into account relativistic corrections to the thermal Stark effect, which acts as the dominant thermal shift of atomic levels, in this part of the paper we give estimates for the effect of finite lifetimes (natural widths) of excited states, see section~\ref{reg}. It is worth noting that the relativistic corrections calculated above are at the level of dynamical thermal contributions, which appear as primary constraints on the experimental accuracy of frequency measurements in atomic clocks due to temperature fluctuations, see, e.g., \cite{Beloy}. By providing estimates for the hydrogen atom, one can assume that the contribution of such effects in neutral atoms is of the same order of magnitude and, therefore, can be used as a rough estimate of relativistic effects in many-electron systems.

To show the importance of finite lifetimes for the energy shift of atomic levels, we turn again to the nonrelativistic limit. The thermal Stark shift can be evaluated using the Sokhotski-Plemelj formula. Then, by regularizing the expression (\ref{Energy shift}) with the Low procedure \cite{Low}, the correction arising from the finite lifetime of excited states can be found as the difference of these two corresponding values. Analytical derivations can be found in \cite{PhysRevA.92.022508}.

Discarding the relativistic corrections, Table~\ref{tab:4} presents the relative contribution of the finite lifetime effect.
\begin{table}[ht]
\centering
\caption{The finite lifetimes effect on the energy shift of atomic levels. The values of relative contribution are given at a temperatures of $20$, $77$, $270$, $300$ and $330$ K for the first four excited $ns(p)$ states of the hydrogen atom.}
\begin{tabular}{|c | c| c | c | c | c|}
\hline
    state & $20$ K & $77$ K  & $270$ K & $300$ K & $330$ K \\
\hline
          
$2s$ & $2.5\times 10^{-3}$ & $-2.6\times 10^{-5}$ &  $-1.5\times 10^{-6}$  & $-1.0\times 10^{-6}$ & $0.0$\\

$2p$ & $2.4\times 10^{-4}$ & $5.5\times 10^{-6}$ &   $0.0$  & -- & --\\

$3s$ & $8.6\times 10^{-4}$ & $1.1\times 10^{-5}$ &  $1.7\times 10^{-7}$ & $0.0$ & --\\

$3p$ & $1.6\times 10^{-2}$ & $-9.1\times 10^{-6}$ &  $0.0$ & -- & --\\
\hline

\end{tabular}

\label{tab:4}
\end{table}
Values indicated as $0.0$ imply an insignificant effect covered by the numerical inaccuracy of the calculations, and hence further calculations are irrelevant.

In particular, from Table~\ref{tab:4} it follows that the effect of finite lifetimes can reach significant magnitude. It is more pronounced at low temperatures for highly excited states. For example, the effect of finite lifetime of levels becomes more pronounced for $3p$ state, where the well-known phenomenon of sign change as a function of temperature occurs, consistent with the polarizability of the atom. The influence of finite lifetimes of atomic levels was first observed in \cite{PhysRevLett.108.043005} as an effect of enhanced thermal friction force. However, at low temperatures, where the effect is considerable, the ac-Stark shift itself is negligible. Therefore, at present, the influence of finite lifetimes in estimating the thermal shift of the atomic energy levels is completely insignificant.
 


\subsection{Thermal Zeeman shift}

In this section, we present numerical results of the BBRZ shift for transitions between hyperfine sublevels for various atoms and ions. The results obtained at $300$ K are collected in Table~\ref{tab:2}, where the first column indicates the selected atomic system, the second column shows the hyperfine transition under consideration. The thermal Zeeman shift is given in the third column in Hz, and its fractional value is specified in the last one.
\begin{table}[ht]
\caption{The BBRZ shift $\Delta \nu^{\mathrm{BBRZ}} \equiv \Delta E_a^{\rm BBRZ} - \Delta E_{a'}^{\rm BBRZ}$ for ground state hyperfine transitions at temperature $T=300$ K.}
\begin{center}
\begin{tabular}{c c c c}
\hline
\hline
System & Transition $a'\rightarrow a$ & $\Delta \nu^{\mathrm{BBRZ}}$, Hz & $\Delta \nu^{\mathrm{BBRZ}}/\nu $\\
\hline
$^1$H            & $F=1\rightarrow F=0$ & $-1.852 \times 10^{-8}$& $-1.304 \times 10^{-17}$\\ 
$^{43}$Ca$^{+}$  & $F=4\rightarrow F=3$ & $-4.206 \times 10^{-8}$& $-1.304 \times 10^{-17}$\\ 
$^{87}$Rb        & $F=2\rightarrow F=1$ & $-8.912 \times 10^{-8}$ & $-1.304 \times 10^{-17}$ \\ 
$^ {87}$Sr$^{+}$  & $F=5\rightarrow F=4$ &$-6.528 \times 10^{-8}$ &  $-1.304 \times 10^{-17}$  \\ 
$^{133}$Cs       & $F=4\rightarrow F=3$ &  $-1.199 \times 10^{-7}$ &$ -1.304 \times 10^{-17}$\\ 
 \hline
\end{tabular}
\end{center}

   \label{tab:2}
\end{table}

From the results in Table~\ref{tab:2}, the value of the thermal Zeeman shift for the ground state hyperfine transition in the hydrogen atom agrees well with the results of the \cite{Itano,Han}. However, it is negligibly small even with respect to the accuracy of the experiment of \cite{Hellwig}. In turn, the fractional shift shown in the last column of the table clearly demonstrates the importance of $\Delta \nu^{\mathrm{BBRZ}}$ for atomic systems used as clocks. The value of the fractional shift is conserved from system to system, varying, however, in absolute value.

The results can be further analyzed in the context of the finite lifetime effect of atomic states. The corresponding relative contributions are given in Table~\ref{tab:2a}. The values are obtained by subtracting the results represented by Eq.~(\ref{BBRZ-2}) from those calculated by the Sokhotski-Plemelj theorem. Hyperfine splitting values are borrowed from \cite{HH-tab}.
\begin{table}[ht]
\caption{The relative contribution of the finite lifetime effect to the BBRZ shift for the hyperfine transitions in the hydrogen atom at different temperatures.}
\centering 
\begin{tabular}{c c c c c}
\hline
\hline
State & $4$ K & $20$ K & $77$ K & $300$ K\\
\hline

$1s_{1/2}^{F=0}$ & $4.7\times 10^{-2}$ & $9.8\times 10^{-3}$ & $2.6\times 10^{-3}$ & $5.8\times 10^{-4}$\\

$2s_{1/2}^{F=0}$ & $0.0$ & -- & -- & --\\

$2p_{1/2}^{F=0}$ & $-1.1\times 10^{-3}$ & $-2.9\times 10^{-4}$ & $-6.4\times 10^{-5}$ & $0.0$\\

$2p_{3/2}^{F=1}$ & $-1.1\times 10^{-3}$ & $-2.2\times 10^{-4}$ & $-4.9\times 10^{-5}$ & $-2.1\times 10^{-5}$\\
 \hline
\end{tabular}
   \label{tab:2a}
\end{table}

Comparing the values collected in Table~\ref{tab:2a} with those in Table~\ref{tab:4}, one can see that the effect of finite lifetimes is more significant for the Zeeman shift. This circumstance is primarily related to the magnitude of the splitting, which falls on the bulk of the Planck distribution, see, e.g., \cite{PhysRevA.110.043108}. However, it is still negligibly small, and its behavior with temperature is approximately the same as for the Stark effect, i.e., its significance decreases with increasing temperature.

Below we also illustrate the contribution of the remaining sum with $n\neq a$ in Eq.~(\ref{Zshift}). This sum arises only with the relativistic correction to the radial wave function, as in the case of the one-photon $2s-1s$ transition probability \cite{Sucher_1978}, i.e., it should be $\alpha^2$ times smaller for the hydrogen atom. Nevertheless, a larger contribution can be expected when summed over the entire spectrum of intermediate states. 
The results of this estimation are collected Table~\ref{tab:2b}.
\begin{table}[ht]
\caption{Contribution of remaining sum for the real part of the expression (\ref{Zshift}), $\delta E_{a}\rightarrow \sum\limits_{n\neq a'}$, for different states in the hydrogen atom at $300$ K. Each second line in each row contains the values for summation over the lower states ($n<a$), whereas the first line contains the higher states ($n>a$). The total contribution is indicated in every third subline and is marked as total. All values are in Hz.}
\begin{center}
\begin{tabular}{ c c c c }
\hline
\hline
State & $\delta E_a $ & State & $\delta E_a$ \\
\hline

$1s_{1/2}$ & $-5.0\times 10^{-12}$ & $3s_{1/2}^{F=0}$ & $-1.7\times 10^{-14}$ \\
           &   $--$ 					  &       & $\,\,\,\,1.0\times 10^{-12}$ \\
total      & $-5.0\times 10^{-12}$ & total & $\,\,\,\,9.9\times 10^{-13}$ \\
\hline

$2s_{1/2}^{F=0}$ & $-1.5\times 10^{-13}$ & $3p_{1/2}^{F=0}$ & $-2.3\times 10^{-15}$\\
                 & $\,\,\,\,3.0\times 10^{-12}$        &         & $\,\,\,\,8.0\times 10^{-14}$\\
total            & $\,\,\,\,2.8\times 10^{-12}$            & total & $\,\,\,\,5.6\times 10^{-14}$\\
\hline
 
$2p_{1/2}^{F=0}$ & $-1.7\times 10^{-13}$ & $3p_{3/2}^{F=1}$ & $-1.4\times 10^{-14}$\\
                 & $--$                    &             & $\,\,\,\,2.2\times 10^{-14}$\\
total            & $-1.7\times 10^{-14}$ & total & $\,\,\,\,8.3\times 10^{-15}$\\
\hline
 
$2p_{3/2}^{F=1}$ & $-6.0\times 10^{-14}$ & $3d_{3/2}^{F=1}$ & $-5.7\times 10^{-15}$ \\
                 & $--$                    &              & $--$ \\
total            & $-6.0\times 10^{-14}$ & total & $-5.7\times 10^{-15}$ \\
 \hline
\end{tabular}
\end{center}

   \label{tab:2b}
\end{table}

In particular, it follows from Table~\ref{tab:2b} that the contribution of the remaining sum is negligible and is always of the order of $10^{-4}$ relative to the leading contribution of the thermal Zeeman shift $n=a'$. Nevertheless, expecting a similar contribution for many-electron systems, the level shifts due to such summation may be at the level of dynamical corrections.

Finally, a detailed study of the Zeeman thermal shift for the most pressing atomic systems has recently been presented in \cite{PhysRevA.110.043108}. According to the analysis presented above, one can indicate that, restoring the values obtained in this work, the influence of relativistic corrections, as well as the account of finite lifetimes of levels, entails only a change of the third-fourth digits of the Zeeman thermal shift. Thus, at the present level of accuracy, the leading order contribution $n=a'$ is sufficient.

\subsection{Imaginary part of the one-loop self-energy correction}

As was shown above, the imaginary part of the one-loop self-energy correction represents the BBR-induced level width $\Gamma^{\beta}_a$, Eq.~(\ref{level width}). Using Sokhotski-Plemelj formula, the corresponding values can be calculated with the results of \cite{PhysRevA.23.2397}), and it is these values that are used in astrophysical studies of radiation transfer, see, e.g., \cite{Seager1999,Seager2000}. Recently, it was shown in \cite{PhysRevA.78.042504} that the ground state of an atom has a finite lifetime due to BBR induced transitions to upper states allowed by the external thermal field. In contrast to the natural level width, the summation over intermediate states in the expression for the induced level width includes upper states \cite{PhysRevA.23.2397,PhysRevA.92.022508}. As a result, at $3000$ K the value of $\Gamma_{1s}^\beta = 1.35\times 10^{-8}$ s$^{-1}$ becomes comparable to the spontaneous one-photon decay of the $2s$ state in the hydrogen atom \cite{PhysRevA.78.042504}. In this section, we analyze the line broadening caused by blackbody radiation, taking into account the finite lifetime of atomic levels.

A rigorous QED theory of the resonance contributions regularization arising in photon scattering on an atom was given in \cite{Low}. This regularization procedure, see \cite{PhysRevA.92.022508}, leads to a modification of the induced width (or partial transition rate at a fixed state $n$ in the sum over the spectrum):
\begin{eqnarray}
\label{gnlb}
\Gamma_a^{\mathrm{ind}+\mathrm{FL}} = \frac{2e^2}{3\pi}\sum\limits_n |(a|\bm{r}|n)|^2 \int\limits_0^\infty d\omega\,\omega^3\,n_\beta(\omega)\times
\\
\nonumber
\left[\frac{\Gamma_{na}}{(E_{na}-\omega)^2+\frac{1}{4}\Gamma_{na}^2}+\frac{\Gamma_{na}}{(E_{na}+\omega)^2+\frac{1}{4}\Gamma_{na}^2}\right],
\end{eqnarray}
where $\Gamma_{na} = \Gamma_n+\Gamma_a$. 

It should be noted that the numerical values in Table III of \cite{PhysRevA.92.022508} are flawed due to an incorrect conversion factor and poor numerical calculation methods. Therefore, we revise the relevant calculations. The results are summarized in Table~\ref{tab3}.
\begin{widetext}
\begin{center}
\begin{table}[ht]
\centering
 \caption{Thermally induced atomic state widths considering finite lifetimes and without. For comparison, each first line shows the values calculated according to the Sokhotski-Plemelj formula ($\Gamma_a^{\mathrm{ind}}$), while each second line shows the results with finite lifetimes ($\Gamma_a^{\mathrm{ind+FL}}$) according to the formula (\ref{gnlb}). All values are in s$^{-1}$.}
\begin{tabular}{c | c | c | c | c | c | c }
\hline
\hline
    state & $77$ K & $270$ K  & $300$ K & $3000$ K & $4000$ K & $30000$ K\\
\hline

$1s$ & $0.0$ & $6.3512\times 10^{-182}$ &  $7.0794\times 10^{-163}$  & $1.3549\times 10^{-8}$ & $2.6180\times 10^{-4}$ & $4.5062\times 10^7$\\
     & $2.9247\times 10^{-11}$ & $4.4228\times 10^{-9}$ &  $6.7414\times 10^{-9}$  & $6.9937\times 10^{-5}$ & $6.5759\times 10^{-4}$ & $4.5063\times 10^7$\\
\hline

$2s$ & $3.6136\times 10^{-6}$ & $1.2674\times 10^{-5}$ &  $1.4083\times 10^{-5}$  & $4.7005\times 10^4$ & $3.0611\times 10^5$ & $9.9573\times 10^7$\\
     & $7.4684\times 10^{-3}$ & $9.0973\times 10^{-2}$ &  $1.1227\times 10^{-1}$  & $4.7017\times 10^4$ & $3.0613\times 10^5$ & $9.9573\times 10^7$\\
\hline

$3s$ & $1.8952\times 10^{-6}$ & $1.0799\times 10^{-5}$ &  $7.8665\times 10^{-5}$  & $9.7311\times 10^5$ & $2.1870\times 10^6$ & $6.3981\times 10^7$\\
     & $8.3118\times 10^{-2}$ & $1.0212$ &  $1.2607$  & $9.7324\times 10^5$ & $2.1872\times 10^6$ & $6.3993\times 10^7$\\
\hline

$4s$ & $1.1296\times 10^{-6}$ & $4.9583$ &  $1.5945\times 10^1$ & $1.6572\times 10^6$ & $2.8751\times 10^6$ & $4.3354\times 10^7$\\
     & $4.0331\times 10^{-1}$ & $9.2339$ &  $2.2067\times 10^1$ & $1.6578\times 10^6$ & $2.8762\times 10^6$ & $4.3415\times 10^7$\\
\hline

$5s$ & $9.5352\times 10^{-6}$ & $5.8097\times 10^2$ &  $1.1964\times 10^3$ & $1.7466\times 10^6$ & $2.7167\times 10^6$ & $3.1376\times 10^7$\\
     & $6.1801\times 10^{-4}$ & $5.8098\times 10^2$ &  $1.1964\times 10^3$ & $1.7466\times 10^6$ & $2.7167\times 10^6$ & $3.1377\times 10^7$\\
\hline
\end{tabular}
    \label{tab3}
\end{table}
\end{center}
\end{widetext}

As can be seen from Table~\ref{tab3}, at room temperature the line broadening caused by thermal radiation, $\Gamma_a^{\mathrm{ind+FL}}$, is significant for low-lying states and is consistent with the "standard"\, value (calculated by the Sokhotski-Plemelj formula) for highly excited states. It can also be seen from Table~\ref{tab3} that at high temperatures the effect of finite lifetimes of atomic states becomes insignificant for any state. This can be directly related to the Lamb shift, since the frequencies of other atomic transitions fall in the tail of the Planck distribution (the value of $n_\beta$ at the resonance frequency is less than unity). At temperatures where the Lamb shift falls on the bulk of the Planck distribution of thermal radiation, regularization by widths becomes necessary. Its role becomes less important as the blackbody radiation peak shifts with increasing temperature.

In addition to the finite lifetime effect for the $1s$ ground state, the results for the $2s$ state attract attention. At room temperature, the contribution induced by blackbody radiation is about $9\%$ of the natural width, $\Gamma_{2s}^{\rm nat} = 8.229$ s$^{-1}$, exceeding the relativistic and QED corrections \cite{JentE1E1}, see also the analysis of linewidth broadening in \cite{Kol}. The contribution, taking into account the effect of finite lifetimes of atomic levels, in the case of the $2s$ state of the hydrogen atom remains dominant up to temperatures of the order of a few thousand Kelvin.

We have also calculated the thermal broadening of the magnetic dipole lines with and without the finite lifetimes of the atomic levels. The results of calculations in the purely quantum mechanical approach for the hydrogen atom are summarized in Table~\ref{tab:6}. As a demonstration, we also present calculation results for several light hydrogen-like ions.
\begin{widetext}
\begin{center}
\begin{table}
\centering
\caption{Partial transition probabilities corresponding to the emission of a magnetic dipole (M1) photon in the hydrogen atom. The transitions are given in the second column, the third column gives the natural level width, the spontaneous probabilities are presented in the fourth column, and the following columns present the results for the induced probabilities due to blackbody radiation and total contributions. All values are given in s$^{-1}$.}

\begin{tabular}{c c  c  c  c c}
\hline
\hline
System & Transition $a'\rightarrow a$ & $\Gamma_{i}^{\rm nat}$ & $W^{\mathrm{spon}}_{if}(\mathrm{M1})$ & $W^{\mathrm{ind}}_{if}(\mathrm{M1})$  & $W^{\mathrm{spon}}_{if}(\mathrm{M1})+W^{\mathrm{ind}}_{if}(\mathrm{M1})$  \\
\hline
H & $1s_{1/2}^{F=1}-1s_{1/2}^{F=0}$ & $2.8689\times 10^{-15}$ & $2.8689\times 10^{-15}$ & $1.2624\times 10^{-11}$ & $1.2627\times 10^{-11}$\\

H & $2s_{1/2}^{F=1}-2s_{1/2}^{F=0}$ & $8.229$ & $5.6040\times 10^{-18}$ & $1.9729\times 10^{-13}$ & $1.9730\times 10^{-13}$\\

H & $2s_{1/2} - 1s_{1/2}$ & $8.229$ & $2.4958\times 10^{-6}$ & $1.1749\times 10^{-4}$ & $1.1999\times 10^{-4}$\\

H & $3s_{1/2} - 1s_{1/2}$ & $6.3169\times 10^6$ & $1.1092\times 10^{-6}$ & $4.3973\times 10^{-5}$ & $4.5082\times 10^{-5}$\\

H & $3s_{1/2} - 2s_{1/2}$ & $6.3169\times 10^6$ & $1.8778\times 10^{-9}$ & $4.8148\times 10^{-7}$ & $4.8336\times 10^{-7}$\\

He$^+$ & $3s_{1/2} - 2s_{1/2}$ & $1.011\times 10^8$ & $1.9233\times 10^{-6}$ & $1.2256\times 10^{-4}$ & $1.2448\times 10^{-4}$\\

Li$^{2+}$ & $3s_{1/2} - 2s_{1/2}$ & $5.117\times 10^8$ & $1.1094\times 10^{-4}$ & $3.1112\times 10^{-3}$ & $3.2221\times 10^{-3}$\\

Ne$^{9+}$ & $3s_{1/2} - 2s_{1/2}$ & $6.318\times 10^{10}$ & $18.9075$ & $39.6399$ & $58.5474$\\
\hline
\end{tabular}
\label{tab:6}
\end{table}
\end{center}
\end{widetext}

The values of the spontaneous emission probabilities presented in Table~\ref{tab:6} are in good agreement with those in \cite{HFS,Popov2017}. It is not surprising that induced probabilities are significantly larger than spontaneous probabilities. This is due to the value of the Planck distribution at the resonance frequency $n_\beta(\omega_0)>1$. Being weakly dependent on the finite lifetimes of atomic levels, nevertheless the effect is observable for the $3s-2s$ transition in the hydrogen atom. Its magnitude is approximately the same for the above systems, giving a relative contribution of about $0.5\,\%$. For all others we found it negligibly small, this is understandable due to the absence of a contribution in the magnetic transitions corresponding to either the Lamb shift or fine structure, see the discussion of the previous table. Note that only partial values are considered here, not the total width. In the latter case, we need to sum over all magnetic transitions, including higher states. Since magnetic transitions are strongly suppressed with respect to dipole transitions, it can be argued that the level widths induced by thermal radiation are formed mainly by dipole transitions to higher states.

The partial probabilities of induced magnetic dipole transitions can also be analyzed on account of the behavior on the nuclear charge $Z$. The lower part of Table~\ref{tab:6} presents the values of the $3s-2s$ transition probabilities in hydrogen and hydrogen-like ions of He, Li, and Ne atoms. This transition is chosen due to the resonance frequency arriving at the bulk of the blackbody distribution. This transition is chosen for the reason that this resonance frequency in the hydrogen atom arrives at the bulk of the blackbody distribution. In particular, for the helium ion it can be seen that at room temperature the partial transition probability caused by BBR still exceeds the spontaneous M1 transition by about two orders of magnitude. The significance decreases with increasing nucleus charge $Z$, where spontaneous magnetic transitions will eventually prevail. The latter can be checked using the results for the hydrogen-like ion Ne$^{9+}$ when the spontaneous and induced probabilities become comparable. 

It should also be noted that this leads to the conclusion that for magnetic transitions the relativistic corrections to emission probabilities are not essential for small $Z$ even at room temperature \cite{Popov2017}: the main contribution is determined by the induced transition rates, whereas the spontaneous one itself represents the leading order correction to them. Thus the effect of finite lifetimes of atomic levels becomes in principle observable for the $2s$ state in the hydrogen atom at room temperature, see, for example, Table~\ref{tab3}. We should also mention that the present accuracy of the experimental determination of atomic level widths is at the level of $10^{-3}$ relative magnitude \cite{Nicholson_2015}. The accuracy of the spectral line width determination at this level of precision made it possible to verify to a large extent the theoretical calculation methods \cite{Martin-PRA}.

Concluding the analysis, numerical results for the thermal Zeeman shift and magnetic line broadening at $300$ K are summarized in Table~\ref{tab:7}, including some values recovered from \cite{PhysRevA.110.043108}. The values of the partial magnetic dipole transition rates were obtained with the use of data in \cite{PhysRevA.110.043108} and can be compared with those in \cite{NIST_ASD}. The given digits are only for direct comparison of the results obtained with and without taking into account the finite lifetime of atomic levels. The level widths are given in s$^{-1}$.
\begin{widetext}
\begin{center}
\begin{table}[ht]
\centering
 \caption{The BBRZ shift, $\Delta \nu^{\rm BBRZ}\equiv\Delta E^{\rm BBRZ}_{a}-\Delta E^{\rm BBRZ}_{a'}$, in Hz for certain atomic clock transitions $a'\rightarrow a$. The first column lists the transition considered in the atomic species, the second column provides the values of the transition energies in cm$^{-1}$, and the third column contains the BBRZ shift in Hz. The following columns BBR-induced width, and BBR-induced width with the account for finite lifetime (FL) of atomic levels.}
\begin{tabular}{ c c  c c  c  c }
\hline
\hline
System & Transition $a'\rightarrow a $& $\Delta E_{a'a}$, cm$^{-1}$ & $\Delta \nu^{\rm BBRZ}$, Hz & $\Gamma^\beta_{a}(\mathrm{M1})$, s$^{-1}$ & $\Gamma^{\beta+\mathrm{FL}}_a(\mathrm{M1})$, s$^{-1}$ \\
\hline

Be& $2s^2(^1S_{0})-2s2p(^3P_{0})$ & $0.6$ & $1.1978\times 10^{-7}$  &  $1.3996\times 10^{-9}$  & $1.3996\times 10^{-9}$  \\

Mg& $3s^2(^1S_{0})-3s3p(^3P_{0})$ & $20$ & $3.8985\times 10^{-6}$  &  $1.4437\times 10^{-6}$  & $1.4437\times 10^{-6}$  \\

Al$^+$& $3s^2(^1S_{0})-3s3p(^3P_{0})$ & $61$ & $1.1201\times 10^{-5}$ & $1.2066\times 10^{-5}$  & $1.2066\times 10^{-5}$  \\

In$^+$& $5s^2(^1S_{0})-5s5p(^3P_{0})$ & $1075$ & $3.5261\times 10^{-5}$ &  $1.3018\times 10^{-4}$  & $1.3018\times 10^{-4}$  \\

Hg$^+$ & $5d^{10}6s(^2S_{1/2})-5d^96s^2(^2D_{5/2})$ & $15041$ & $-4.4990\times 10^{-7}$ & $1.7335\times 10^{-30}$  & $2.3920\times 10^{-14}$  \\
\hline
\end{tabular}

    \label{tab:7}
\end{table}
\end{center}
\end{widetext}

The values given in Table~\ref{tab:7} demonstrate the insignificance of the effect of finite lifetimes of atomic levels in the thermal Zeeman effect. This is primarily due to the small magnitude of the BBRZ shift itself. As in the case of the electric dipole transition, the modification of the induced transition probabilities due to finite lifetimes of atomic levels becomes essential only at large values of the corresponding widths and frequencies of the essentially off bulk Planck distribution.

\section{Consclusion}

This study presents a thorough theory and accompanying analysis of energy level shifts in an atom subjected to a blackbody radiation field. The calculations presented mainly for the hydrogen atom are intended to demonstrate the theory of this paper in the simplest possible way. In particular, the widely discussed dynamical Stark shift induced by blackbody radiation and the induced level widths are evaluated in the framework of QED theory employing the thermal one-loop formalism. The numerical results of the thermal Stark shift for different states at laboratory temperatures are summarized in Table~\ref{Table1}, the results for decay rates see in Tables~\ref{tab3}, \ref{tab:6}, \ref{tab:7}. 

The TQED approach reveals additional thermal effects associated with relativistic corrections. In particular, a careful analysis of the decomposition of the one-loop self-energy operator in the case of finite temperatures allows one to identify the relativistic correction to the thermal Stark effect (thermal shift of the leading order), see Eq.~(\ref{BBRr}). Also there is a thermal correction arising from the relativistic contribution from the wave function in the nonrelativistic approximation, see Eq.~(\ref{Lc4}). The corresponding numerical values are collected in Tables~\ref{tab:rem} and \ref{tab:3}, respectively, and their cumulative contribution is presented in the Table~\ref{tab:3a}. Although depending on the considered states in the hydrogen atom, these corrections still exceed the thermal Zeeman shift (see Table~\ref{tab:2} for the latter) and are comparable to the dynamical corrections to the dc-Stark shift. Thus, the relativistic corrections to the thermal shifts are important for analyzing the error (at least) of the determined transition frequencies. The importance of the obtained thermal corrections of relativistic nature is also emphasized by the fact that using temperatures at which the Stark effect is canceled out, these effects can appear to dominate in the temperature fluctuations of the transition frequency.

In addition, a detailed study of the effect of finite lifetime (level width) of atomic levels on the Stark thermal shift is carried out in this paper, see also \cite{PhysRevLett.108.043005,PhysRevA.92.022508}. The impact of the corresponding modification on the Stark shift can be expressed similarly as in Eq.~(\ref{BBRZ-2}), see \cite{PhysRevA.92.022508} for details. A numerical analysis of the effect as a function of temperature and for different states is presented in section~\ref{lifetimes}, see Tables~\ref{tab:4}, \ref{tab:2a}. In contrast to the recently considered in \cite{PhysRevLett.108.043005} enhancement off BBR-friction force due to the inclusion of atomic-level widths a similar effect for the Stark shift is found to be negligible for the hydrogen atom at room temperature.

The present study also provides a rigorous derivation of the thermal Zeeman shift within the one-loop formalism of the TQED approach. The numerical results of the BBRZ shift for hyperfine transitions in several atoms and ions are given in Table~\ref{tab:2} (see also Table~\ref{tab:7} for other transitions). A comparative analysis taking into account the finite lifetimes for the Zeeman thermal shift can be done by using Table~\ref{tab:2a}. We did not consider relativistic corrections to the Zeeman thermal shift (as in the case of BBRS), since it is itself a relativistic correction according to the TQED formalism. However, we have presented rough estimates for the sum over intermediate states, usually discarded in calculations of the BBRZ effect, see Table~\ref{tab:2b}. This "residual"\, sum arises taking into account relativistic corrections \cite{Sucher_1978} and, as expected, gives negligibly small contribution at the present level of experimental accuracy. However, the relative contribution for the considered states of the hydrogen atom is approximately $10^{-4}$ with respect to the BBRZ shift.

The paper also presents analytical expressions and numerical calculations for the widths and partial probabilities of transitions induced by thermal radiation, see Tables~\ref{tab3}, \ref{tab:6} and \ref{tab:7}. The corresponding quantities arise naturally within the one-loop TQED formalism as the imaginary part of the thermal correction to the self-energy of the bound electron. The influence of finite lifetimes of atomic levels on these quantities has been studied in detail. As for the real part, the influence is more pronounced at low temperatures and for low states. It is worth noting that for partial probabilities corresponding to magnetic dipole transitions, it is the induced quantities that are dominant (not spontaneous) and, therefore, the effect of finite lifetimes is most important in this case. This circumstance should be extremely important when comparing theoretical calculations of the matrix elements and corresponding experimental results \cite{Nicholson_2015}.

Furthermore, we note that within the formalism under the study, thermal quadrupole interaction also arises, see Eq.~(\ref{BBRQ}), as well as the following higher multipoles can be obtained as a result of the decomposition of the one-loop self-energy operator. We have not presented a numerical analysis of this value due to its smallness. However, this effect is considered for the corresponding transitions in atoms, see, e.g. \cite{Safronova2010,PhysRevA.83.012503}. Within the framework of the used formalism we also derived the thermal diamagnetic offset of atomic levels, see subsection~\ref{di}. The obtained analytical expressions (\ref{di.2}) and (\ref{di.3}) enable us to calculate the corresponding thermal shift for any atomic systems. We emphasize that all the effects considered in this study arise within the framework of one one-loop formalism of the QED theory at a finite temperature.

Finally, it should be noted that the thermal propagator (the second term in Eq.~(\ref{phpr})) leads to a fundamental difference from the phenomenological approach usually performed within the framework of QM theory, see, e.g., the discussion in \cite{PhysRevA.109.012816}. The latter leads to multiplication of the dynamical polarizability by the Planck distribution function followed by integration over frequency, see Eq.~(\ref{farley'sformula}). In this case the absence of additional relativistic contributions in the quantum mechanical approach becomes understandable, whereas the one-loop formalism is more consistent and rigorous.

\section*{Acknowledgements}
The work of the author T.Z. was supported by a grant from the Foundation for the Advancement of Theoretical Physics and Mathematics "BASIS"\, No.~23-1-3-31-1. The work of authors J.R.L.J., A.B., P.K., and D.S. was supported by a grant from the Foundation for the Advancement of Theoretical Physics and Mathematics "BASIS"\, No.~25-1-2-18-1.

\bibliographystyle{apsrev4-1}
\bibliography{References}

\appendix

\setcounter{equation}{0}
\setcounter{figure}{0}
\setcounter{table}{0}
\renewcommand{\thefigure}{\thesection\arabic{figure}} 
\renewcommand{\thetable}{\thesection\arabic{table}}

\section{Derivation of thermal Zeeman and quadrupole shifts}
\label{ApB}
\subsection{Derivation of thermal Zeeman shift}
\label{ApBadd}
To obtain the thermal Zeeman shift of the atomic level in the framework of QED theory, one should consider a correction of the next order with respect to the dipole approximation in the one-loop self-energy operator described by the Feynman graph in Fig.~\ref{fig:1}. The corresponding matrix element is
\begin{eqnarray}
\label{Z1}
 \langle  a n| (\bm{\alpha}_1 \bm{\alpha}_2)r^2_{12}  | n a \rangle = 
 \\
 \nonumber
 \langle  a n| (\bm{\alpha}_1 \bm{\alpha}_2) (r^2_1+r^2_2-2(\bm{r}_1\bm{r}_2)) | n a \rangle. \,\,\,\,\,
\end{eqnarray}
Further it is convenient to use the relation
\begin{eqnarray}
\label{Z2}
([\bm{r}_1\times\bm{\alpha}_1]\cdot[\bm{r}_2\times\bm{\alpha}_2]) =  
\\
\nonumber
(\bm{\alpha}_1 \bm{\alpha}_2) (\bm{r}_1\bm{r}_2) - 
(\bm{r}_1\bm{\alpha}_2)(\bm{r}_2\bm{\alpha}_1),
\end{eqnarray}
which enables one to distinguish the magnetic moment operator, $\bm{\mu}$, with the help of definition, see \cite{PhysRevA.56.R2499},
\begin{eqnarray}
\label{Z3}
\bm{\mu} = \frac{e}{2}[\bm{r}\times\bm{\alpha}].
\end{eqnarray}
In the expressions above, the dot denotes the scalar product, and the vector product is denoted by a cross.

Then it should be taken into account that the operator in Eq.~(\ref{Z1}) also includes the quadrupole momentum, see, e.g., \cite{Solovyev_2010} and references therein. To derive the Zeeman thermal shift, we use only half of the last contribution in the right-hand side of Eq.~(\ref{Z1}). The result can be written as
\begin{eqnarray}
\label{Z4}
\langle  a n| (\bm{\alpha}_1 \bm{\alpha}_2)r^2_{12}  | n a \rangle = 
\\
\nonumber
-\langle  a n| ([\bm{r}_1\times\bm{\alpha}_1]\cdot[\bm{r}_2\times\bm{\alpha}_2]) | n a \rangle +\,\,\,\,\,
\\
\nonumber
+\langle  a n| (\bm{\alpha}_1 \bm{\alpha}_2) (r^2_1+r^2_2) | n a \rangle
\\
\nonumber
- \langle  a n| (\bm{r}_1 \bm{\alpha}_2) (\bm{r}_2 \bm{\alpha}_1) - (\bm{\alpha}_1 \bm{\alpha}_2) (\bm{r}_1 \bm{r}_2) | n a \rangle.
\end{eqnarray}

Then for the first term in the right-hand side of Eq.~(\ref{Z4}) using Eq.~(\ref{Z3}), the energy shift (\ref{relc}) reduces to
\begin{eqnarray}
\label{Z5}
\Delta E_a^{\mathrm{BBRZ}} = -\frac{2}{3\pi}\sum\limits_n\langle a n|   (\bm{\mu}_1\bm{\mu}_2) |n a \rangle \int\limits^{\infty}_0 d\omega\, n_{\beta}(\omega)\,\omega^3 \,\,\,\,\
  \\
  \nonumber
  \times
  \left( \frac{1}{\varepsilon_a-\varepsilon_n + \omega + i 0} +  \frac{1}{\varepsilon_a-\varepsilon_n - \omega + i 0}   \right),
\end{eqnarray}
where the $1/2$ factor in the definition of $\bm{\mu}$ quadruples the total coefficient in the expression (\ref{relc}), the electron charge $e$ is also included in the definition of $\bm{\mu}$, see Eq.~(\ref{Z3}). Note also that it is completely relativistic expression (wave functions in matrix elements as well as energies in denominators correspond to Dirac quantities).

For this purpose one should use the relations
\begin{eqnarray}
\label{Z6}
[r_i,p_j] &=& \mathrm{i} \delta_{ij},
\nonumber
\\
(\bm{\sigma}\bm{a})\bm{\sigma}  &=& \bm{a} + \mathrm{i} [\bm{\sigma} \times \bm{a}],
\\
\nonumber
\bm{\sigma}(\bm{\sigma}\bm{a})  &=& \bm{a} + \mathrm{i} [\bm{a} \times  \bm{\sigma}].
\end{eqnarray}
Here $\bm{p}$ is the momentum operator, $\bm{\sigma}$ is the Pauli matrix, and the last two expressions are written for an arbitrary vector $\bm{a}$. The component of the vector product $[\bm{r}\times\bm{\alpha}]_i$ can be quoted as $[\bm{r}\times \bm{\alpha}]_i = \epsilon_{ijk} r_j\alpha_k$ ($\epsilon_{ijk}$ is the Levi-Civita symbol).

Setting aside for simplicity the strictness of writing the expressions, the matrix element $\langle  a| \mu_i  | n \rangle$ can be written as
\begin{eqnarray}
\label{Z7}
\langle  a| \mu_i  | n \rangle = \frac{|e|}{2}\epsilon_{ijk}\langle  a|r_j\alpha_k  | n \rangle \rightarrow
\nonumber
\\
\begin{pmatrix}
\varphi_a^* & \chi_a^*
\end{pmatrix} \begin{pmatrix}
0 & r_{j}\sigma_{k}\\
r_{j}\sigma_{k} & 0
\end{pmatrix} \begin{pmatrix}
\varphi_n \\
 \chi_n
\end{pmatrix} \approx 
\\
\nonumber
\varphi_a^* \left( r_{j}\sigma_{k} \frac{(\bm{\sigma} \bm{p})}{2m_e} + \frac{(\bm{\sigma} \bm{p})}{2m_e} r_{j}\sigma_{k}    \right) \varphi_n,
\end{eqnarray}
where we have used the coupling of the small and large components of the Dirac wave function $\chi\approx \frac{(\bm{\sigma} \bm{p})}{2m_e}\varphi$.

Then, applying the last two ratios in Eq.~(\ref{Z6}), one can find
\begin{eqnarray}
\label{Z8}
\begin{pmatrix}
\varphi_a^* & \chi_a^*
\end{pmatrix} \begin{pmatrix}
0 & r_{j}\sigma_{k}\\
r_{j}\sigma_{k} & 0
\end{pmatrix} \begin{pmatrix}
\varphi_n \\
 \chi_n
\end{pmatrix} \approx \qquad
\\
\nonumber
\frac{1}{2m_e}  \varphi_a^* \left( r_jp_k + ir_j[\bm{p} \times \bm{\sigma}]_k +
 p_k r_j + i[\bm{\sigma} \times \bm{p}]_k r_j     \right)\varphi_n.
\end{eqnarray}
Using the commutation relation (the first line in (\ref{Z6})) for the last two summands and the index summation properties for the product of Levy-Civita tensors, the final expression can be represented as 
\begin{eqnarray}
\label{Z9}
\langle  a| \mu_i  | n \rangle \approx \frac{|e|}{2m_e}( a|[\bm{r}\times \bm{p}]_i + \sigma_i|n),
\end{eqnarray}
where Schr\"odinger wave functions are implied in the matrix element envelopes. 
It is easy to see that Eq.~(\ref{Z9}) reduces to the matrix element from the well-known expression for the nonrelativistic magnetic dipole moment operator, which is $\bm{\mu}^{\mathrm{nr}} = \mu_{\rm B}(\bm{l}+2\bm{s})$. Here $\bm{l}$ is the orbital momentum of the electron, $\bm{s} = \bm{\sigma}/2$ is the spin of the electron, and $\mu_{\rm B}=|e|/2m_e$ is the Bohr magneton.
Thus, the expression (\ref{Z5}) is completely identical to that obtained in the framework of quantum mechanics theory, see, e.g., \cite{PhysRevA.110.043108}.

\subsection{Derivation of thermal quadripole interaction}
\label{ApBquadr}
To construct the quadrupole moment, one should consider the remaining terms arising from the last two summands in the right-hand side of Eq.~(\ref{Z4}), i.e. $\langle a n| (\bm{r}_1 \bm{\alpha}_2) (\bm{r}_2 \bm{\alpha}_1) + (\bm{\alpha}_1 \bm{\alpha}_2) (\bm{r}_1 \bm{r}_2) | n a \rangle$. Let us first transform this matrix element to the symmetric form:
\begin{widetext}
\begin{eqnarray}
\label{2Z1}
\langle a n| (\bm{\alpha}_1 \bm{\alpha}_2) (\bm{r}_1 \bm{r}_2) + (\bm{r}_1 \bm{\alpha}_2) (\bm{r}_2 \bm{\alpha}_1) | n a \rangle =
\langle a n| \alpha_{1i}\alpha_{2i} r_{1j}r_{2j} + \alpha_{1i}\alpha_{2j} r_{1j}r_{2i} | n a \rangle
\end{eqnarray}
and, therefore,
\begin{eqnarray}
\label{2Z1a}
\frac{1}{2}\langle a n| \alpha_{1i}\alpha_{2i} r_{1j}r_{2j} + \alpha_{1i}\alpha_{2j} r_{1j}r_{2i} | n a \rangle + 
\frac{1}{2}\langle a n| \alpha_{1j}\alpha_{2j} r_{1i}r_{2i} + \alpha_{1j}\alpha_{2i} r_{1i}r_{2j} | n a \rangle = 
\\
\nonumber
\frac{1}{2}\langle an | (\alpha_{1i}r_{1j}+\alpha_{1j}r_{1i})(\alpha_{2i}r_{2j}+\alpha_{2j}r_{2i}) | n a\rangle
\equiv
\frac{1}{2}\langle a| \alpha_{1i}r_{1j}+\alpha_{1j}r_{1i}| n\rangle\langle n|\alpha_{2i}r_{2j}+\alpha_{2j}r_{2i}| a\rangle.
\end{eqnarray}
\end{widetext}
The Einstein summation convention is assumed here, implying summation over the set of indexed terms in a formula.

Then, applying the nonrelativistic limit as in the previous section, one can obtain
\begin{eqnarray}
\label{2Z2}
\langle a| \alpha_{1i}r_{1j} + \alpha_{1j}r_{1i}|n\rangle 
\approx 
\\
\nonumber
\frac{1}{m_e}(a| r_{1i}p_{1j}+r_{1j}p_{1i}-\mathrm{i}\delta_{ij}|n),
\\
\nonumber
\langle n| \alpha_{1i}r_{1j} + \alpha_{1j}r_{1i}|a\rangle 
\approx 
\\
\nonumber
\frac{1}{m_e}(n| r_{2i}p_{2j}+r_{2j}p_{2i}-\mathrm{i}\delta_{ij}|a).
\end{eqnarray}
The contribution of the Kronecker delta in the matrix elements of Eq.~(\ref{2Z2}) can be excluded from further consideration since it leads to the equivalence of states $a$ and $n$. The latter gives zero in the final result because of the proportionality to the energy difference $\omega_{an} = E_a-E_n$ (see below).

Further transformations involve the use of the commutation relation: $(n| r_i p_j + r_j p_i |a) = \mathrm{i}m_e(n| [H,r_i r_j] |a) = \mathrm{i}m_e(E_n-E_a)(n| r_i r_j |a)$, see Eq. (4.295) in Ref. \cite{JentAdkins}. Identification of an operator of type $r_i r_j$ corresponds to the part of the quadrupole moment. To obtain the remaining part in $3r_i r_j - r^2\delta_{ij}$, we need to consider the next order of $\alpha Z$ decomposition of $\sin$ in the expression (\ref{Energy shift}), which is given by the contribution $\omega^5r^4_{12}/120$. Thus, we can write for the real part of the energy shift (\ref{relc}) without considering the Zeeman thermal shift:
\begin{widetext}
\begin{eqnarray}
\label{2Z3}
\Delta E_a^{\rm rel-} = \frac{e^2}{6\pi}\sum\limits_n\int\limits_0^\infty d\omega\, \omega^3 n_\beta(\omega) 
\frac{2\varepsilon_{an}}{\varepsilon_{an}^2-\omega^2}
\langle a n| (\bm{\alpha}_1 \bm{\alpha}_2)(r_1^2+r_2^2) + \frac{\omega^2}{20}(r_1^4+r_2^4-4(\bm{r}_1 \bm{r}_2)(r_1^2+r_2^2)) | n a\rangle  -\qquad
\\
\nonumber
\frac{e^2}{6\pi}\sum\limits_n\int\limits_0^\infty d\omega\, \omega^3 n_\beta(\omega)
\frac{\omega_{an}^3}{\omega_{an}^2-\omega^2}
(a|r_{1i}r_{1j}|n)(n|r_{2i}r_{2j}|a) + 
\frac{e^2}{6\pi}\sum\limits_n\int\limits_0^\infty d\omega\, \omega^3 n_\beta(\omega)
\frac{2\varepsilon_{an}}{\varepsilon_{an}^2-\omega^2}
\langle a n| \frac{\omega^2}{5}(\bm{r}_1\bm{r}_2)^2+\frac{\omega^2}{10}r_1^2r_2^2 |n a\rangle
\end{eqnarray}
with corresponding notations for relativistic (nonrelativistic) energies and wave functions. The integrals are implied in the sense of the principal value according to Sokhotski-Plemelj theorem. Herewith for matrix elements depending only on radius-vectors, we can immediately pass to the leading contribution, replacing relativistic energies and wave functions by their nonrelativistic analogs (by large components of Dirac functions in particular).

The second integral in Eq.~(\ref{2Z3}) can be further transformed by adding and subtracting a similar contribution, but proportional to $\omega_{an}\omega^2$, so that the energy denominator is reduced. Then, leaving the first term in Eq.~(\ref{2Z3}) unchanged, we arrive at
\begin{eqnarray}
\label{2Z4}
\Delta E_a^{\rm rel-} = \text{first} -
\frac{e^2}{6\pi}\sum\limits_n\int\limits_0^\infty d\omega\, \omega^3 n_\beta(\omega)
\omega_{an} (a n|(\bm{r}_1 \bm{r}_2)^2|n a) + 
\\
\nonumber
-\frac{e^2}{6\pi}\frac{3}{5}\sum\limits_n\int\limits_0^\infty d\omega\, n_\beta(\omega)
\frac{\omega_{an}\omega^5}{\omega_{an}^2-\omega^2}
( a n| (\bm{r}_1\bm{r}_2)^2 |n a) +
\frac{e^2}{6\pi}\frac{1}{5}\sum\limits_n\int\limits_0^\infty d\omega\, n_\beta(\omega)
\frac{\omega_{an}\omega^5}{\omega_{an}^2-\omega^2}(a n| r_1^2r_2^2 |n a)
\end{eqnarray}
\end{widetext}
Taking into account that $(3r_{1i} r_{1j}-r^2_1\delta_{ij})(3r_{2i} r_{2j}-r^2_1\delta_{ij}) \equiv \mathcal{Q}^{(1)}_{ij} \mathcal{Q}^{(2)}_{ij} = 9(\bm{r}_1\bm{r}_2)^2-3r_1^2r_2^2$, one can find the thermal energy shift caused by the quadrupole interaction:
\begin{eqnarray}
\label{2Z5}
\Delta E_a^{\mathcal{Q}} = - \frac{e^2}{90\pi}\sum\limits_n\int\limits_0^\infty d\omega\, 
\frac{\omega_{an} \omega^5 n_\beta}{\omega_{an}^2-\omega^2}
(an| \mathcal{Q}^{(1)}_{ij} \mathcal{Q}^{(2)}_{ij} |na).\,\,\,\,\,\,\,
\end{eqnarray}

The structure of the shift $\Delta E_a^{\mathcal{Q}}$ is the same as for the Stark or Zeeman thermal shifts, but it is $\beta^{-2}$ times smaller (excluding the prefactor). For the case of $\omega\ll \omega_{an}$, the expression (\ref{2Z5}) can be decomposed into a Taylor series, which immediately gives the $T^6$ dependence, i.e., as dynamical corrections to the static Stark shift induced by blackbody radiation. Finaly, the expression (\ref{relc}) can be reduced to
\begin{widetext}
\begin{eqnarray}
\label{2Z6}
\Delta E_a^{\rm rel} \approx   \Delta E_a^{\mathcal{B}} + \Delta E_a^{\mathcal{Q}} + \frac{e^2}{6\pi}\sum\limits_n\int\limits_0^\infty d\omega\, \omega^3 n_\beta(\omega) 
\frac{2\omega_{an}}{\omega_{an}^2-\omega^2}
( a n| (\bm{\alpha}_1 \bm{\alpha}_2)(r_1^2+r_2^2) | n a) - \qquad
\\
\nonumber
\frac{e^2}{6\pi}\sum\limits_n\int\limits_0^\infty d\omega\, \omega^3 n_\beta(\omega)
\omega_{an} (a n|(\bm{r}_1 \bm{r}_2)^2|n a) + 
\frac{e^2}{60\pi}\sum\limits_n\int\limits_0^\infty d\omega\,  \omega^5 n_\beta(\omega)
\frac{\omega_{an}}{\omega_{an}^2-\omega^2}
(a n|r_1^4+r_2^4-4(\bm{r}_1 \bm{r}_2)(r_1^2+r_2^2)|n a)
\\\nonumber
=\Delta E_a^{\mathcal{B}} + \Delta E_a^{\mathcal{Q}} +
\Delta E_a^{\rm rem}.
\end{eqnarray}
\end{widetext}

Keeping in mind also that the part of $\Delta E_a^{\rm rel}$ may refer to the higher multipoles \cite{Porsev} and the correction to the matrix element of the magnetic dipole \cite{Beigman_1971,Drake_M1,Sucher_1978}, we discard further consideration of terms proportional to $\omega^5$ in the integrands. In the main text, the calculations are performed for levels of hyperfine and fine structure, i.e., when $n= n_a$ in $\Delta E_a^{\mathcal{B}}$. We also give results for summation over $n\neq n_a$ states in the Zemman shift in a fully relativistic way, which can be done by simple replacement of the radial integral with Schr\"odinger wave functions by its relativistic analog $\int\,dr\, r^2(f_{n_a\kappa_{a}}g_{n\kappa_{n}}+f_{n\kappa_{n}}g_{n_a\kappa_{a}})$, where $f$, $g$ are the large and small radial components of the Dirac wave function, respectively, and $\kappa_{n}=(j_{n}+\frac{1}{2})(-1)^{j_{n}+l_{n}+s}$, see \cite{Puchkov2009}.

\subsection{The remaining terms in Eq.~(\ref{2Z6})}
\label{rem}

The derivation of the thermal magneto-dipole and electric quadrupole interactions from the one-loop correction to the self-energy of the bound electron left out the terms proportional to $\omega^3$. The importance of these contributions to the frequency shift should be verified, since they are temperature coincident with the dc-Stark thermal effect (dominant thermal shift). Carrying out standard estimates for hydrogen-like ions, the third and fourth terms in Eq.~(\ref{2Z6}) (proportional to $\omega^3$) can be valued as $\sim(k_{\rm B}T)^4/(m\alpha Z^2)$ r.u. (or $\alpha^5(k_{\rm B}T)^4/(m Z^2)$ in a.u.), i.e., they are $\alpha^2$ times smaller for the hydrogen atom relative to the Stark shift and of the same order as the BBRZ shift.

Let us denote $\Delta E_a^{\rm rem}$ in Eq.~(\ref{2Z6}) as $J_3+J_4+J_5$ according to its constituent terms. We first consider a simplification of the third contribution, which is of the form:
\begin{eqnarray}
\label{3Z1}
J_{3}=
\frac{e^2}{6\pi}\sum\limits_n\int\limits_0^\infty d\omega\, \omega^3 n_\beta(\omega) 
\frac{2\varepsilon_{an}}{\varepsilon_{an}^2-\omega^2}
\\\nonumber
\times
\langle a n| (\bm{\alpha}_1 \bm{\alpha}_2)(r_1^2+r_2^2) | n a\rangle 
\end{eqnarray}
Here, as before, the integral over the frequency $\omega$ is implied in the sense of the principal value obtained by the Sokhotski-Plemelj formula. 
Going to the nonrelativistic limit, in the leading order we have
\begin{eqnarray}
\label{3Z2}
J_{3}=
\frac{2e^2}{3\pi}\sum\limits_{nk}\int\limits_0^\infty d\omega\, \omega^3 n_\beta(\omega) 
\frac{\omega_{an}^2\omega_{ak}}{\omega_{an}^2-\omega^2}
\\\nonumber
\times
( a | r_i| k) ( k | r^2 | n) ( n | r_i | a),
\end{eqnarray}
where the projector $\sum_{k}|k\rangle \langle k|$ and commutator relation  $-\mathrm{i}m_e(n|[r_i,H]|a) = (n|p_i|a)$ were used. With this transformation, the contribution of the negative spectrum vanishes.

Further simplification of $J_{3}$ contribution corresponds to adding and subtracting the term 
$\omega_{ak} \omega^2 / (\omega_{an}^2 - \omega^2)$ to Eq.~(\ref{3Z2}) with the same matrix elements. 
Then, we obtain
\begin{eqnarray}
J_{3} = \frac{2e^2}{3\pi} \sum\limits_{nk} \int\limits_0^\infty d\omega\, 
    \omega^3 n_\beta(\omega) \omega_{ak} 
    ( a | r_i | k) ( k | r^2 | n) ( n | r_i | a)
\nonumber \\
     + \frac{2e^2}{3\pi} \sum\limits_{nk} \int\limits_0^\infty d\omega\, 
    \frac{\omega_{ak} \omega^5 n_\beta(\omega) }{\omega_{an}^2 - \omega^2} 
( a | r_i | k) ( k | r^2 | n) ( n | r_i | a).\,\,\,\,\,\,\,
    \label{3Z4}
\end{eqnarray}
The first contribution in (\ref{3Z4}) can be summed over $n$ subject to the completeness condition, and then the summation index $k$ can be replaced by $n$. 

Next, the fifth contribution in Eq.~(\ref{2Z6}) can be reduced to
\begin{eqnarray}
J_{5} = -\frac{e^2}{15\pi} \sum\limits_n \int\limits_0^\infty d\omega\, n_\beta(\omega)\omega^5
\frac{\omega_{an} }{\omega_{an}^2 - \omega^2}
\\
\nonumber
\times (a n | (\bm{r}_1 \bm{r}_2)(r_1^2 + r_2^2) | n a).
\end{eqnarray}
Here we have taken into account that $(a n| r_1^4+r_2^4| na) = 2\delta_{an}(a|r^4|n)$ that gives zero via $\omega_{an}$ in the numerator of the last term in Eq.~(\ref{2Z6}).
Then, substituting the results for $J_{3}$ and $J_{5}$ into the contribution $\Delta E_a^{\rm rem}$ of the expression (\ref{2Z6}), we obtain
\begin{widetext}
\begin{eqnarray}
\label{3Z5}
\Delta E_a^{\rm rem} =  
\frac{e^2}{6\pi}\sum\limits_{n}\int\limits_0^\infty d\omega\, \omega^3 n_\beta(\omega) 
\omega_{an} ( a n | 2 (\bm{r}_1 \bm{r}_2) (r_1^2+r_2^2) | n a) - 
\frac{e^2}{6\pi}\sum\limits_n\int\limits_0^\infty d\omega\, \omega^3 n_\beta(\omega)
\omega_{an} (a n|(\bm{r}_1 \bm{r}_2)^2|n a) +\qquad
\\
\nonumber
\frac{2e^2}{3\pi}\sum\limits_{nk}\int\limits_0^\infty d\omega\, \omega^5 n_\beta(\omega) \frac{\omega_{ak}}{\omega_{an}^2-\omega^2} ( a | r_i| k) ( k | r^2 | n) ( n | r_i | a) - 
\frac{e^2}{15\pi}\sum\limits_n\int\limits_0^\infty d\omega\, \omega^5n_\beta(\omega)
\frac{\omega_{an} }{\omega_{an}^2-\omega^2}
(a n|(\bm{r}_1 \bm{r}_2)(r_1^2+r_2^2)|n a).
\end{eqnarray}
\end{widetext}

Discarding the higher order contributions (the last two terms $\sim T^6$ ), the result for the relativistic correction is
\begin{eqnarray}
\label{3Z6}
\Delta E_a^{\rm rem} =  
\frac{e^2}{6\pi}\sum\limits_{n}\int\limits_0^\infty d\omega\, \omega^3 n_\beta(\omega) 
\omega_{an} 
\\
\nonumber
\times
( a n | 4 (\bm{r}_1 \bm{r}_2) r_1^2 - (\bm{r}_1 \bm{r}_2)^2| n a).
\end{eqnarray}
Turning back to the  relations $-\mathrm{i}m_e(n|[r_j,H]|a) = (n|p_j|a)$ and $\mathrm{i}m_e(n|[H, r_ir_j]|a) =(n|r_ip_j+r_jp_i|a)$, the final expression for the correction $\Delta E_a^{\rm rem}$ can be found as
\begin{eqnarray}
\label{3Z7}
\Delta E_a^{\rm rem} =   
\frac{e^2}{3\pi}\frac{\pi^4(k_{\rm B} T)^4}{15}
( a |  r^2(\bm{r}\cdot \bm{\nabla})| a),
\end{eqnarray}
The results of nonrelativistic calculations (using the quantum theory of angular momentum \cite{varsh}) are collected in the main text of the article in Table~\ref{tab:rem}.

\section{Diamagnetic contribution: negative spectrum}
\label{dia}

It is well known that passing from the Dirac equation to the nonrelativistic limit in the framework of the Pauli approximation, consideration of the generalized momentum ($\bm{\pi}= \bm{p}+\frac{e}{c}\bm{A}$) leads to the contribution quadratic on the vector-potential $\bm{A}$ (external field) \cite{JentAdkins}. The arising relativistic correction is proportional to $1/c^2$ and in the case of a homogeneous magnetic field represents the diamagnetic interaction of the electron. A sequential calculation within the quantum mechanics theory shows that this correction corresponds to the operator $ \mathcal{B}^2 \bm{r}^2$. In present part of the work we show that within QED approach employed throughout our analysis, the same correction can be obtained from the one-loop self energy correction as a contribution of negative spectrum omitted upto now from the consideration. 

To derive the diamagnetic correction one can start from the expressions (\ref{Energy shift}) and (\ref{Ir12}). For the real part the negative spectrum contribution can be written as
\begin{eqnarray}
\label{dia.1}
 \Delta E_a^{(-)} = \frac{e^2}{\pi} \sum_{n^{(-)}} \left[  \frac{1-\bm{\alpha}_1\bm{\alpha}_2}{r_{12}} I^{\beta}_{n^{(-)} a}(r_{12})\right]_{an^{(-)} n^{(-)} a},\,\,\,\,
 \\
 \nonumber
I^{\beta}_{n^{(-)} a}(r_{12}) = \int\limits^{\infty}_0 d\omega\, n_{\beta}(\omega) \sum\limits_{\pm}\frac{\sin\omega r_{12}}{\varepsilon_a-\varepsilon_{n^{(-)}} \pm \omega + \mathrm{i}\,0},
\end{eqnarray}
where the states from negative spectrum are denoted by $n^{(-)}$. 

In further one can take into account that $\varepsilon_a-\varepsilon_{n^{(-)}}\approx 2mc^2$, the Dirac energies for the positive and negative spectra are separated by the energy gap equal to the two electron rest mass. Since the frequency of the BBR is given by the Planck's distribution function one can neglect by $\omega$ in the energy denominator of Eq.~(\ref{dia.1}) in respect to $2mc^2$ and the sum $\sum_\pm$ leads to the factor $2$. Then, expanding the $\sin$ into the Taylor's series,
\begin{eqnarray}
\label{dia.2}
I^{\beta}_{n^{(-)} a}(r_{12}) \approx \frac{1}{m_e c^2}\int\limits^{\infty}_0 d\omega\, n_{\beta}(\omega)\left[\omega r_{12}-\frac{\omega^3}{6}r_{12}^3+\dots
\right],
\end{eqnarray}
with $r_{12}^2=r_1^2+r_2^2-2\bm{r}_1\bm{r}_2$. Here and below we explicitly write $m_e$ and $c$ for clarity.

Substituting Eq.~(\ref{dia.2}) into the expression (\ref{dia.1}) (first line) leads to the following contributions:
\begin{eqnarray}
\label{dia.3}
 \Delta E_a^{(-)}\approx 
\frac{e^2}{\pi m_e c^2}\sum_{n^{(-)}}\int\limits^{\infty}_0 d\omega\, n_{\beta}(\omega)\omega\, \delta_{a n^{(-)}} - \qquad
\\
\nonumber
\frac{e^2}{\pi m_e c^2}\sum_{n^{(-)}}\int\limits^{\infty}_0 d\omega\, n_{\beta}(\omega)\langle an^{(-)}|\omega \bm{\alpha}_1\bm{\alpha}_2+ \frac{\omega^3}{3}\bm{r}_1\bm{r}_2|n^{(-)} a\rangle 
\\
\nonumber
-\frac{e^2}{3\pi m_e c^2}\sum_{n^{(-)}}\int\limits^{\infty}_0 d\omega\, n_{\beta}(\omega)\omega^3\langle a|\bm{r}^2|n^{(-)} \rangle\, \delta_{a n^{(-)}}+\dots
\end{eqnarray}
The first and third terms in Eq.~(\ref{dia.3}) are zero due to the orthogonality condition of the wave functions from the positive and negative Dirac spectra and, therefore, only the second should be evaluated additionally.

Taking into account the completeness condition for negative spectrum functions (which can be imposed in the non-relativistic limit \cite{Landau}), we found
\begin{eqnarray}
\label{dia.4}
 \Delta E_a^{(-)}\approx - \frac{e^2}{\pi m_e c^2}\int\limits^{\infty}_0 d\omega\, n_{\beta}(\omega)\omega\,\langle a| \bm{\alpha}^2+ \frac{\omega^2}{3}\bm{r}^2| a\rangle
\end{eqnarray}
The first summand here can be simplified by the relation $\bm{\alpha}_i\bm{\alpha}_j = \delta_{ij}+\mathrm{i}\varepsilon_{ijk}\bm{\Sigma}_k$ \cite{LabKlim}, which gives a factor of $3$ for matching indices $i,j$ (the Levy-Civita symbol vanishes in this case). Thus, the term with $\bm{\alpha}$ matrices gives a state-independent contribution. This constant, however, exactly cancels the analogous state-independent part of the Stark thermal shift, when forming a cubic dependence on $\omega$ (see Eq.~(141) in \cite{SOLOVYEV2020168128}). In the simplest case, it is sufficient to merely discard all such contributions. 

Finally, the remaining term in Eq.~(\ref{dia.4}) gives precisely the diamagnetic thermal shift, which we denote hereafter as $\Delta E_a^{\rm dia}$ (in units where $m_e=c=1$):
\begin{eqnarray}
\label{dia.5}
\Delta E_a^{\rm dia} = \frac{\pi^3 e^2}{45\beta^4}\langle a | r^2| a\rangle.
\end{eqnarray}
For the hydrogen atom the matrix element is $\langle a | r^2| a\rangle = \frac{n_a^2}{2}(5n_a^2+1-3l_a(l_a+1))$, where $n_a$ is the principal number and $l_a$ is the orbital momentum for the arbitrary state $a$. To obtain the magnitude of the thermal diamagnetic shift the coefficient in Eq.~(\ref{dia.5}) can be expressed in terms of room temperature as
\begin{eqnarray}
\label{dia.6}
\frac{\pi^3 e^2}{45\beta^4}\rightarrow 7.6425\times 10^{-8} \left[\frac{T}{300\, \text{K}}\right]^4 \text{ Hz.}
\end{eqnarray}

\end{document}